%% file: strmixeff.tex
\documentclass[aps,prd,preprint,a4paper,showpacs]{revtex4}
\usepackage{graphicx,natbib}
\usepackage{amsmath}
\newcommand{\bi}[1]{\mbox{\boldmath ${#1}$}}

\begin{document}
\title{Effects to  Scalar Meson Decays of Strong 
Mixing  between Low and High Mass Scalar Mesons}
\vspace*{1cm}
\author{T. Teshima}
\email{teshima@isc.chubu.ac.jp}
\author{I. Kitamura}
\author{N. Morisita}
\affiliation{Department of Applied Physics,  Chubu University, Kasugai 
487-8501, Japan}
\begin{abstract}
We analyze the mass spectroscopy of low and high mass scalar mesons and get 
the result that the coupling strengths of the mixing between low and high 
mass scalar mesons are very strong and the strengths of mixing for $I=1,\ 
1/2$ scalar mesons and those of $I=0$ scalar mesons are almost same. Next, 
we analyze the decay widths and decay ratios of these mesons and get the 
results that the coupling constants $A'$ for $I=1,\ 1/2$ which represents the 
coupling of high mass scalar meson $N'$ $\to$ two pseudoscalar mesons $PP$ are 
almost same as the coupling $A'$ for the $I=0$. On the other hand, the coupling 
constant $A$ for $I=1,\ I=1/2$ which represents the low mass scalar meson $N$ 
$\to PP$ are far from the coupling constant $A$ for $I=0$. We consider a 
resolution for this discrepancy. Coupling constant $A''$ for glueball $G \to 
PP$ is smaller than the coupling $A'$. $\theta_P$ is $40^\circ \sim 50^\circ$.
\end{abstract}
\preprint{CU-TP/03-05}
\maketitle
\section{Introduction}
In recent re-analyses of $\pi\pi$ scattering phase sift and production 
processes, the existence of scalar mesons $\sigma(500)$ (we call $f_0(500)$ 
hereafter) have been confirmed \cite{sigma}, and in the analyses of $K\pi$ 
scattering phase sifts and production processes, the existence of 
$\kappa(900)$ have been reported \cite{kappa}. The $f_0(500)$ was considered 
to be a chiral partner of the $\pi$ meson as a Nambu-Goldstone boson 
\cite{chiral} and the $f_0(500)$ and $\kappa(900)$ are considered  to construct 
the low mass scalar nonet together with  $a(980)$ and $f_0(980)$. 
This nonet has been considered to be a chiral partner of the ground state 
pseudoscalar nonet in connection with the linear sigma model \cite{linear}.  
Many authors analyzed the nonet using the $K\bar{K}$ molecule model 
\cite{molecule} or $qq\bar{q}\bar{q}$ model \cite{jaffe} rather than the 
linear sigma model in order to explain the $s\bar{s}$ rich character of the 
$f_0(980)$ which degenerate to $a_0(980)$.  On the other hand, high mass scalar 
mesons $a(1450)$, $K_0^*(1430)$, $f_0(1370)$ and $f_0(1710)$ are considered 
to construct the high mass scalar nonet. This nonet is considered as the 
ordinary $L=1\ q\bar{q}$ scalar nonet.
\par
We assume a strong mixing (inter-mixing) between low mass and high mass scalar 
nonets to explain the fact that the high mass $L=1\ q\bar{q}$ scalar nonet are 
so high compared to other $L=1\ q\bar{q}$ $1^{++}$ and $2^{++}$ mesons 
\cite{schechter,teshima}. This assumption is supported by the viewpoint in 
which the $a_0(980)$ and $f_0(980)$ contain the four-quark, two-quark and
meson-meson contents \cite{achasov}.  
Furthermore $f_0(1500)$ is considered to be a glueball candidate \cite
{glueball}. Thus, this glueball mixes with $I=0$ $L=1\ q\bar{q}$ 
scalar mesons, and furthermore mixes with low mass scalar $f_0(980)$ through 
inter-mixing \cite{achasov,teshima}.  We thus analyzed the overall mixing 
among low mass $qq\bar{q}\bar{q}$ scalar nonet and $L=1$ $q\bar{q}$ scalar 
nonet and glueball \cite{teshima}. We have obtained the result that the 
inter-mixing is very strong and the mixing parameters $\lambda_{01}^a$, 
$\lambda_{01}^K$ and $\lambda_{01}$ producing the inter-mixing in $I=1$, $I=1/2$ 
and $I=0$ mesons respectively are almost same \cite{teshima}. 
\par 
If there exists the strong mixing between low and high mass scalar mesons, the 
decay processes of low mass scalar mesons may be affected by the high mass scalar 
mesons and glueball, and conversely the decay processes of high mass scalar mesons 
may be affected by the low mass scalar mesons. Black et al. \cite{schechter} 
estimated the decay coupling constants $A$ for low mass scalar mesons ($N$)-
pseudoscalar meson($P$)-pseudoscalar meson($P$) interaction and the coupling 
constant $A'$ for high mass scalar($N'$)-$PP$ interaction considering the mixing 
between $I=1$ and $I=1/2$ low and high mass scalar mesons. We also analyze 
these decay coupling constants $A$ and $A'$ and further the coupling 
constant $A''$ for glueball($G$)-($PP$) interaction considering the mixing 
among $I=0$ low and high mass scalar mesons. The values of $(A,\ A')$ estimated 
in $I=1$ and $I=1/2$ meson decay analyses are $\sim(0.1\ -3)$, while these 
estimated in $I=0$ meson decays are $\sim(-4,\ -2)$ for the case where 
$f_0(1710)$ is considered as glueball and $\sim(-2.9,\ -2.3)$ for the case 
where $f_0(1500)$ is considered as glueball. There is a large discrepancy between 
the values of $A$ in $I=1, 1/2$ and $I=0$ cases. We will discuss about a resolution 
for this discrepancy. 
\section{Mixing between Low and High Mass Scalar Mesons}
In this section, we briefly review the mixing among the low mass scalar, high 
mass scalar and glueball discussed in our previous work \cite{teshima}.
\subsection{ Structure of low mass scalar mesons}
For the structures of the low mass scalar mesons, there considered two 
possibilities. One is the chiral partner of the pseudoscalar nonet 
\cite{chiral} and the other is the $qq\bar{q}\bar{q}$ \cite{jaffe} or 
$M\overline{M}$ molecule \cite{molecule}. We assume 
the $qq\bar{q}\bar{q}$ structure because of the degeneracy between $a_0$ and 
$f_0(980)$ which has large $s\bar{s}$ character. This is understood readily 
from the flavor contents in $qq\bar{q}\bar{q}$ mesons: 
\begin{equation}
    \begin{array}{ccc}
    \bar{s}\bar{d}us,\ \frac{1}{2}(\bar{s}\bar{d}ds-\bar{s}\bar{u}us),\ 
    \bar{s}\bar{u}ds&\Longleftrightarrow&a_0^+,\ a_0^0,\ a_0^-\\
    \bar{s}\bar{d}ud,\ \bar{s}\bar{u}ud,\ \bar{u}\bar{d}us,\ \bar{u}\bar{d}ds
    &\Longleftrightarrow&\kappa^+,\ \kappa^0,\ \overline{\kappa^0},\ \kappa^-\\
    \frac{1}{2}(\bar{s}\bar{d}ds+\bar{s}\bar{u}us)&\Longleftrightarrow&f_N
    \sim f_0(980)\\
    \bar{u}\bar{d}ud&\Longleftrightarrow&f_S\sim f_0(500) 
    \end{array}
\end{equation}
\par 
The masses of $I=0$ $f_0(980)$ and $f_0(500)$ mesons are represented by the 
masses of $a_0(980)$, $\kappa(900)$ and mixing mass parameter $\lambda_0$, which 
causes the mixing (intra-mixing) between $f_0(980)$ and $f_0(500)$ and describes 
the interaction strength of OZI rule suppression graph shown in Fig. 1. 
Diagonalizing the mass matrix 
\begin{equation}
\left(
\begin{array}{cc}
m^2_{a_0}+2\lambda_0&\sqrt{2}\lambda_0\\
\sqrt{2}\lambda_0&2m^2_\kappa-m^2_{a_0}+\lambda_0
\end{array}\right)
\end{equation}
and using the relation $m_s>m_{u,d}$, we can get the desired spectrum, 
$$m^2_{f_0(980)}\approx m^2_{a_0(980)}>m^2_\kappa>m^2_{f_0(500)}.$$
\subsection{Inter-mixing between $I=1,\ 1/2$ low mass scalar mesons 
and $I=1,\ 1/2$ high mass scalar mesons}
The inter-mixing interaction is caused by the graph shown in Fig. 2, which 
represents the OZI rule allowed interaction. 
\begin{eqnarray}
L_{\rm int}&=&-\lambda_{01}\epsilon^{abc}\epsilon_{def}N^d_a
    N'^e_b\delta^f_c=\lambda_{01}[a_0^+{a'}_0^-+a_0^-{a'}_0^++a_0^0{a'}_0^0
    +\kappa^+K_0^{*-}      \nonumber\\
    &&+\kappa^-K_0^{*+}+\kappa^0K_0^{*0}+\bar{\kappa}^+
    \bar{K}_0^{*-}-\sqrt{2}f_Nf'_N-f_Sf'_N-\sqrt{2}f_Nf'_S],
\end{eqnarray}
where $N'^a_b=q_b\bar{q}^a$. The strength of this $\lambda_{01}$ is considered 
to be very large because of the OZI rule allowed interaction. 
\begin{figure}
\vspace{0.5cm}
\begin{center}
\input{fig1.tex}\hspace{2cm} \input{fig2.tex} 
\vspace{0.5cm}\\
Fig. 1\hspace{7.5cm}Fig. 2
\end{center}
\end{figure}
\par
We estimate the strength of the inter-mixing parameter $\lambda_{01}$. First, 
we estimate that for $I=1$ $a_0(1450)$ and $a_0(980)$ mixing case.
We estimate the masses before mixing as
\begin{equation}
m_{\overline{a_0(980)}}=1271\pm31{\rm MeV},\ \ m_{\overline{a_0(1450)}}=
1236\pm20{\rm MeV}  
\end{equation}
from the relation $m^2(2^{++})-m^2(1^{++})=2(m^2(1^{++})-m^2(0^{++}))$ 
resulted from the $L\cdot S$ force. The mass value $1271\pm31{\rm MeV}$ 
of the $\overline{a_0(980)}$ and $\overline{f_0(980)}$ before mixing is almost 
same to the values $1275{\rm MeV}$ for $a_0(980)$ and $1282{\rm MeV}$ for 
$f_0(980)$ estimated in the constituent 4 quarks model \cite{tetra}.   
Diagonalizing the mass matrix    
\begin{equation}
\left(\begin{array}{cc}
    m_{\overline{a_0(980)}} & \lambda_{01}^a\\
    \lambda_{01}^a & m_{\overline{a_0(1450)}}
    \end{array}\right)
\end{equation}
and taking the eigenvalues of masses
\begin{equation}
m_{a_0(980)}=984.8\pm1.4{\rm MeV},\ \ m_{a_0(1450)}=1474\pm19{\rm MeV},
\end{equation}
we can get the result
\begin{equation}
\lambda_{01}^a=0.600\pm0.028{\rm GeV^2}, \ \  \ \ {\rm mixing\ angle\ }
\theta_{a}=47.1\pm3.5^\circ. 
\end{equation}
Next, we estimate the strength $\lambda_{01}$ for $I=1/2$ $\kappa(900)$ and 
$K^*_0(1430)$ mixing case. Using the masses before mixing and after mixing, 
\begin{equation}
\begin{array}{l}
m_{\overline{\kappa(900)}}=1047\pm62{\rm MeV},\ \ m_{\overline{K^*_0(1430)}}=
1307\pm11{\rm MeV}, \\
m_{\kappa(900)}=900\pm70{\rm MeV},\ \ m_{K^*_0(1430)}=1412\pm6{\rm MeV}. 
\end{array}
\end{equation}
we get the results
\begin{equation}
\lambda_{01}^K=0.507\pm84{\rm GeV^2},\ \ \ \  
{\rm mixing\ angle\ }\theta_{K}=29.5\pm15.5^\circ.
\end{equation}
It is confirmed that these coupling strengths are large and $\lambda_{01}^K$ 
is as strong strength as $\lambda_{01}^a$.
\subsection{Inter-mixing between $I=0$ low and $I=0$ high mass scalar mesons 
and glueball}
Intra-mixing between $I=0, L=1\ q\bar{q}$ scalar mesons and glueball are 
expressed by the matrix as 
\begin{equation}
\left(\begin{array}{ccc}
m^2_{a_0}+2\lambda_1&\sqrt{2}\lambda_1&\sqrt{2}\lambda_G\\
\sqrt{2}\lambda_1&2m^2_K-m^2_{a_0}+\lambda_1&\lambda_G\\ 
\sqrt{2}\lambda_G&\lambda_G&\lambda_{GG}
\end{array}\right).
\end{equation}
$\lambda_1$ is the term of the OZI-rule suppression graph for $q\bar{q}$
shown in Fig. 3. 
\begin{figure}
\vspace{0.5cm}
\begin{center}
\input {fig3}\\
Fig. 3
\end{center}
\end{figure}
$\lambda_{G}$ is the transition between $q\bar{q}$ and glueball $gg$ showed 
in Fig. 4 (a) and $\lambda_{GG}$ is the pure glueball mass shown in Fig. 4 (b).
\begin{figure}
\begin{center}
\input{fig4}
\vspace{0.5cm}\\
Fig. 4
\end{center}
\end{figure}
\par
We analyze the inter- and intra-mixing among $I=0$ low mass and high 
mass scalar mesons and glueball expressed by the overall mixing mass matrix 
as
\begin{equation}
\left(\begin{array}{ccccc}
    m^2_{N}+2\lambda_0&\sqrt{2}\lambda_0&\lambda_{01}&\sqrt{2}\lambda_{01}
    &0\\
    \sqrt{2}\lambda_0&m^2_{S}+\lambda_0&\sqrt{2}\lambda_{01}
    &0&0\\
    \lambda_{01}&\sqrt{2}\lambda_{01}&m^2_{N'}+2\lambda_1&\sqrt{2}\lambda_1&
    \sqrt{2}\lambda_{G}\\
    \sqrt{2}\lambda_{01}&0&\sqrt{2}\lambda_1&m^2_{S'}+\lambda_1&
    \lambda_{G}\\
    0&0&\sqrt{2}\lambda_{G}&\lambda_{G}&\lambda_{GG}
    \end{array}\right).
\end{equation}
Using the input mass values (unit:{\rm GeV})
\begin{equation}
\begin{array} {l}   
    m_{N}=1.271\pm0.031, \ m_{S}=0.760\pm0.179, \  m_{N'}=1.236\pm0.02,\\ 
    m_{S'}=1.374\pm0.003, \ m_{f_0(980)}=0.980\pm0.010, \ 
    m_{f_0(500)}=0.500\pm0.100,\\
    m_{f_0(1370)}=1.350\pm0.150,\ m_{f_0(1710)}=1.715\pm0.007, \ 
    m_{f_0(1500)}=1.500\pm0.010,
\end{array}
\end{equation}
we get the result for the case in which $f_0(1500)$ is assumed as glueball.
\begin{eqnarray}
&&\hspace{1cm}\lambda_{01}=0.53\pm0.04{\rm GeV^2},\ \lambda_0=0.03
\pm0.04{\rm GeV^2},\ \ 
\lambda_1=0.07\pm0.05{\rm GeV^2},\nonumber\\
&&\hspace{1cm}\lambda_G=0.23\pm0.06{\rm GeV^2},\ \ \lambda_{GG}=
(1.53\pm0.03)^2{\rm GeV^2},
\nonumber\\
&&\hspace{4cm}\left(\begin{array}{c}
    f_0(980)\\
    f_0(500)\\
    f_0(1370)\\
    f_0(1500)\\
    f_0(1710)
    \end{array}\right)=[R_{f_0(M)I}]\left(\begin{array}{c}
    f_N\\
    f_S\\
    f_{N'}\\
    f_{S'}\\
    f_G
    \end{array}\right),\\    
&&\hspace{1cm}  [R_{f_0(M)I}]=\nonumber\\
&&  \left(\begin{array}{ccccc}
     0.720\pm0.060&-0.389\pm0.096&-0.148\pm0.111&-0.558\pm0.041&0.145\pm0.048\\
     0.234\pm0.093&0.789\pm0.080&-0.525\pm0.080&-0.102\pm0.059&0.108\pm0.035\\
     0.048\pm0.077&0.433\pm0.062&0.683\pm0.039&-0.482\pm0.044&-0.275\pm0.054\\
     -0.416\pm0.119&0.013\pm0.039&0.059\pm0.060&-0.361\pm0.122& 0.812\pm0.103\\
     0.532\pm0.101&0.168\pm0.036&0.459\pm0.049&0.508\pm0.037& 0.453\pm0.172
     \end{array}\right).\nonumber
\end{eqnarray}
Next, we study the case in which $f_0(1710)$ is assumed as glueball. This case was 
not analyzed in our early work \cite{teshima}.
\begin{eqnarray}
&&\hspace{1cm}\lambda_{01}=0.44\pm0.04{\rm GeV^2},\ \lambda_0=0.02
\pm0.05{\rm GeV^2}, \ 
\lambda_1=-0.08\pm0.05{\rm GeV^2},\nonumber\\
&&\hspace{1cm}\lambda_G=0.28\pm0.06{\rm GeV^2}, \ \lambda_{GG}=
(1.64\pm0.03)^2{\rm GeV^2}
, \\
&&\hspace{1cm}  [R_{f_0(M)I}]=\nonumber\\
&&\left(\begin{array}{ccccc}
     0.635\pm0.084&-0.511\pm0.106&-0.210\pm0.160&-0.524\pm0.030&0.147\pm0.074\\
     0.243\pm0.112&0.723\pm0.114&-0.584\pm0.099&-0.174\pm0.077&0.123\pm0.056\\
     0.210\pm0.064&0.424\pm0.066&0.716\pm0.065&-0.482\pm0.088&-0.187\pm0.062\\
     0.651\pm0.051&0.033\pm0.040&0.052\pm0.093&0.599\pm0.100&-0.426\pm0.096\\        0.255\pm0.085&0.086\pm0.031&0.282\pm0.057&0.306\pm0.071& 0.858\pm0.069
     \end{array}\right).\nonumber
\end{eqnarray}
The characters of the mixing parameters of these $I=0$ scalar mesons are 
described as follows; (1) $f_0(980)$ contains $f_N=(\bar{s}\bar{d}ds+
\bar{s}\bar{u}us)/\sqrt{2}$ and $f_{S'}=\bar{s}s$ components 
about $70\sim80\%$, (2) $f_0(500)$ contains $f_S=\bar{u}\bar{d}ud$ and $f_{N'}
=(\bar{u}u+\bar{d}d)/\sqrt{2}$ components 
about $90\%$, (3) $f_0(1370)$ contains $f_N$ and $f_{S'}$ components about 
$70\%$, (4) $f_0(1500)$ contains $f_G$ component about $70\%$ in $f_0(1500)$ 
glueball case, and $f_0(1710)$ contains $f_G$ component about $70\%$ in 
$f_0(1710)$ glueball case.
\section{Decay processes of scalar mesons and glueball}
\noindent
In this section, we analyze the decay processes of low mass scalar mesons 
($N$) decaying to two pseudoscalar mesons ($PP$), high mass scalar mesons 
($N'$) decaying to $PP$ and pure glueball ($G$) decaying to ($PP$). 
We use the following interactions for $NPP$, $N'PP$ and 
$GPP$ coupling with coupling constants $A$, $A'$ and $A''$, respectively, 
\begin{equation}
L_I=A\varepsilon^{abc}\varepsilon_{def}N^d_a\partial^\mu\phi^e_b
\partial_\mu\phi^f_c+
A'N'^b_a\{\partial^\mu\phi^c_b,\ \partial_\mu\phi^a_c\}+
A''G\{\partial^\mu\phi^b_a,\ \partial_\mu\phi^a_b\}.
\end{equation}
These interactions are represented graphically by the diagrams in fig.~5. 
Although interactions as ${\rm Tr}(N\partial_\mu\phi){\rm Tr}(\partial^\mu\phi)$ 
and ${\rm Tr}(N'\partial_\mu\phi){\rm Tr}(\partial^\mu\phi)$ other than those 
represented by Eq.~(15) may exist \cite{schechter}, these 
interactions violate the $OZI$ rule and are considered to be small compared to 
the interactions in Eq.~(15). 
\vspace{0.2cm}\\
\begin{center}
\input{fig51}\hspace{1cm}\input{fig52}\hspace{1cm}\input{fig53}\\
Fig.5
\end{center}
\subsection{$a_0(980)$, $a_0(1450)$ and $K_0^*(1450)$ meson decays}
We define the coupling constants $\gamma_{a_0K\overline{K}}$ etc. in the 
following expression, 
\begin{eqnarray}
L_I&=&\gamma_{a_0K\overline{K}}\frac{1}{\sqrt{2}}\partial_\mu\overline{K}
\bi{\tau}\bi{\cdot}\bi{a_0}\partial^{\mu}K
+\gamma_{a'_0K\overline{K}}\frac{1}{\sqrt{2}}\partial_\mu\overline{K}\bi{\tau}
\bi{\cdot}\bi{a'_0}\partial^{\mu}K
+\gamma_{a_0\pi\eta}\bi{a_0\cdot}\partial_\mu\bi{\pi}\partial^{\mu}\eta
+\gamma_{a'_0\pi\eta}\bi{a'_0\cdot}\partial_\mu\bi{\pi}\partial^{\mu}\eta
\nonumber\\
&+&\gamma_{a_0\pi\eta'}\bi{a_0\cdot}\partial_\mu\bi{\pi}\partial^{\mu}\eta'
+\gamma_{a'_0\pi\eta'}\bi{a'_0\cdot}\partial_\mu\bi{\pi}\partial^{\mu}\eta'
+\gamma_{\kappa K\pi}(\frac{1}{\sqrt{2}}\partial_\mu\overline{K}
\bi{\tau\cdot}\partial^{\mu}\bi{\pi}\kappa+H.C.)\nonumber\\
&+&\gamma_{K^*K\pi}(\frac{1}{\sqrt{2}}\partial_\mu\overline{K}
\bi{\tau\cdot}\partial^{\mu}\bi{\pi}K^*+H.C.)
+\gamma_{\kappa K\eta}(\overline{\kappa}\partial_\mu K\partial^{\mu}\eta
+H.C.)
+\gamma_{K^* K\eta}(\overline{K^*}\partial_\mu K\partial^{\mu}\eta
+H.C.)\nonumber\\
&+&\gamma_{\kappa K\eta'}(\overline{\kappa}\partial_\mu K\partial^{\mu}\eta'
+H.C.)
+\gamma_{K^* K\eta'}(\overline{K^*}\partial_\mu K\partial^{\mu}\eta'
+H.C.),
\end{eqnarray}
where fields $a_0$ represents the low mass $I=1$ scalar mesons and $a_0'$ 
the high mass $I=1$ scalar mesons. Then the  coupling constants for $I=1$ and 
$1/2$ meson decays are, by using Eq.~(15), expressed  as
\begin{eqnarray}
\gamma_{a_0(980)K\overline{K}}&=&2(A \cos\theta_a-A' \sin\theta_a),\nonumber\\
\gamma_{a_0(980)\pi\eta}&=&2(A \cos\theta_a\sin\theta_P-\sqrt{2}A' 
\sin\theta_a\cos\theta_P),\nonumber\\
\gamma_{a_0(1450)K\overline{K}}&=&2(A \sin\theta_a+A' \cos\theta_a),\nonumber\\
\gamma_{a_0(1450)\pi\eta}&=&2(A \sin\theta_a\sin\theta_P+\sqrt{2}A' 
\cos\theta_a\cos\theta_P),\nonumber\\
\gamma_{a_0(1450)\pi\eta'}&=&2(-A \sin\theta_a\cos\theta_P+\sqrt{2}A' 
\cos\theta_a\sin\theta_P),\nonumber\\
\gamma_{\kappa^*(900)\pi K}&=&2(A \cos\theta_K-A' \sin\theta_K),\nonumber\\
\gamma_{K_0^*(1430)\pi K}&=&2(A \sin\theta_K+A' \cos\theta_K),
\end{eqnarray}
where $\theta_P$ is $\eta$-$\eta'$ mixing angle and related to the 
traditional octet-singlet mixing angle $\theta_{0\mbox{-}8}$ as $\theta_P=
\theta_{0\mbox{-}8}+54.7^\circ$. Decay widths of these mesons are expressed 
by using the coupling constants $\gamma_{a_0(980)K\bar{K}}$ etc. as
\begin{eqnarray}
\Gamma(a_0(M)\to K(m_1)+\overline{K}(m_2))&=&\frac{\gamma^2_{a_0(M)K
\overline{K}}}{32\pi}\frac{q_{Mm_1m_2}}{m_{a_0(M)}^2} m_{Mm_1m_2}^4 \ ,
\nonumber \\
\Gamma(a_0(M)\to \pi(m_1)+\eta(m_2))&=&\frac{\gamma^2_{a_0(M)\pi\eta}}{32\pi}
\frac{q_{Mm_1m_2}}{m_{a_0(M)}^2} m_{Mm_1m_2}^4 \ ,
\nonumber\\
\Gamma(a_0(M)\to \pi(m_1)+\eta'(m_2))&=&\frac{\gamma^2_{a_0(M)\pi\eta'}}{32\pi}
\frac{q_{Mm_1m_2}}{m_{a_0(M)}^2} m_{Mm_1m_2}^4 \ ,
\nonumber\\
\Gamma(K_0^*(M)\to \pi(m_1)+ K(m_2))&=&\frac32\frac{\gamma^2_{K_0^*(M)\pi K}}
{32\pi}\frac{q_{Mm_1m_2}}{m_{K_0^*(M)}^2} m_{Mm_1m_2}^4 \ .
\end{eqnarray}
Here $q_{M m_1m_2}$ and $m_{M m_1m_2}$ are defined as 
\begin{eqnarray}
&&q_{M m_1m_2}=\sqrt{(\frac{M^2+m_2^2-m_1^2}{2M})^2-m_2^2}\ ,\nonumber\\
&&m_{M m_1m_2}=\sqrt{M^2-m_1^2-m_2^2}\ ,\nonumber
\end{eqnarray}
and for the case $M\approx m_1+m_2$, we use the next formula for $q_{M m_1m_2}$,
\begin{equation}
q_{M m_1m_2}={\rm Re}\frac{1}{\sqrt{2\pi}\Gamma_M}\int^{M+\infty}_
{M-\infty}e^{-\frac{(m-M)^2}{2\Gamma_M^2}}\times 
\sqrt{(\frac{m^2+m_2^2-m_1^2}{2m})^2-m_2^2}\ dm,
\end{equation}
where $\Gamma_M$ is the decay width of particle with mass $M$. This procedure 
is similar to that of the first article in \cite{sigma}.
\par
We used the data for these decay processes cited in PDG \cite{pdg} and those 
are listed in second column of Table I. Using these data, we estimated the 
allowed values for $A$ and $A'$ in the $\chi^2\le5.348$ corresponding to the 
50\% C.L. on degree of freedom 6. For the mixing 
angles $\theta_a$ and $\theta_K$, we use the results of Eqs. (7) and (9); 
$\theta_a=(47.1\pm3.5)^\circ$ and $\theta_K=(29.5\pm 15.5)^\circ$.  
\begin{table}
\caption{Experimental data and best fit values for various decay widths and 
decay ratios. Best fit values are obtained for $A=0.22,\ 
A'=-3.13,\ \theta_P=49.0^\circ$.}
\begin{center}
\begin{tabular}{|c|c|c|} \hline\hline
Decay width and ratio&Experimental data&Best fit value\\ \hline
$\Gamma(a_0(980)\to {\rm all}(\pi\eta+K\bar{K}))$&$75\pm25{\rm MeV}$&
$36{\rm MeV}$\\ \hline
$\Gamma(a_0(980)\to K\bar{K})/\Gamma(a_0(980)\to \pi\eta)$&$0.177\pm0.024$
&$0.156$\\ \hline
$\Gamma(a_0(1450)\to {\rm all}(\pi\eta+\pi\eta'+K\bar{K}))$&$265\pm13
{\rm MeV}$&$266{\rm MeV}$\\ \hline
$\Gamma(a_0(1450)\to K\bar{K})/\Gamma(a_0(1450)\to \pi\eta)$&$0.88\pm0.23$
&$0.80$\\ \hline
$\Gamma(a_0(1450)\to \pi\eta')/\Gamma(a_0(1450)\to \pi\eta)$&$0.35\pm0.16$
&$0.49$\\ \hline
$\Gamma(K_0^*(1430)\to \pi K)$&$273\pm44{\rm MeV}$&$303{\rm MeV}$\\ \hline
\end{tabular}
\end{center}
\end{table}
Estimated values of $A$, $A'$ and $\theta_P$ are 
\begin{equation}
\begin{array}{l}
A=0.10\pm0.24,\ A'=-3.03\pm0.2,\ \theta_P=49.0^\circ\pm3.0^\circ.\\ 
\end{array}
\end{equation}
We show the best fit values for decay widths and decay ratios on $A=0.22,\ 
A'=-3.13,\ \theta_P=49.0^\circ$ in third column of 
Table I. From this result, one finds that the estimated total width of 
$a_0(980)$ is rather small than experimental width. This may be caused 
from the treatment in which we used the Eq.~(19) to estimate the decay momentum 
$q_{Mm_1m_2}$ when $M\sim m_1+m_2$ for the decay of $a_0(980)$. 
\subsection{$f_0(980)$, $f_0(1370)$,\ $f_0(1500)$ and $f_0(1710)$ 
{\rm meson\ decays}}
If we define the coupling constants $\gamma_{f_0(M)\pi\pi}$ etc. in the 
following expression, 
\begin{eqnarray}
L_I&=&\gamma_{f_0(M)\pi\pi}\frac{1}{2}f_0(M)\partial_\mu\bi{\pi}\bi{\cdot}
\partial^{\mu}\bi{\pi}
+\gamma_{f_0(M)K\overline{K}}f_0(M)\partial_\mu\overline{K}\partial^{\mu}K
+\gamma_{f_0(M)\eta\eta}f_0(M)\partial_\mu\eta\partial^{\mu}\eta
\nonumber\\
&+&\gamma_{f_0(M)\eta\eta'}f_0(M)\partial_\mu\eta\partial^{\mu}\eta'
+\gamma_{f_0(M)\eta'\eta'}f_0(M)\partial_\mu\eta'\partial^{\mu}\eta',
\end{eqnarray}
then the coupling constants $\gamma_{f_0(M)\pi\pi}$ etc. for $f_0(M)$  $(M=980, 
1370, 1500,\ 1710)$ are expressed from the interaction for $NPP$, 
$N'PP$ and $GPP$ coupling represented in Eq.~(15) as 
\begin{eqnarray}
\gamma_{f_0(M)\pi\pi}&=&2(-A R_{f_0{(M)}S}+\sqrt{2}A' R_{f_0{(M)}N'}
+2A'' R_{f_0{(M)}G}),\nonumber\\
\gamma_{f_0(M)K\bar{K}}&=&\sqrt{2}(-A R_{f_0{(M)}N}+A' R_{f_0{(M)}N'}
+\sqrt{2}A' R_{f_0{(M)}S'}+2\sqrt{2} A'' R_{f_0{(M)}G}),\nonumber\\
\gamma_{f_0(M)\eta\eta}&=&2(-A R_{f_0{(M)}N}\cos\theta_P\sin\theta_P
+\frac12A R_{f_0{(M)}S}\cos^2\theta_P\nonumber\\
&+&\frac{1}{\sqrt{2}}A' R_{f_0{(M)}N'}\cos^2\theta_P
+A' R_{f_0{(M)}S'}\sin^2\theta_P+A'' R_{f_0{(M)}G}),\nonumber\\
\gamma_{f_0(M)\eta\eta'}&=&2(A R_{f_0{(M)}N}\cos2\theta_P
+\frac12A R_{f_0{(M)}S}\sin2\theta_P\nonumber\\
&+&\frac{1}{\sqrt{2}}A' R_{f_0{(M)}N'}\sin2\theta_P
-A' R_{f_0{(M)}S'}\sin2\theta_P),\nonumber\\
\gamma_{f_0(M)\eta'\eta'}&=&2(A R_{f_0{(M)}N}\cos\theta_P\sin\theta_P
+\frac12A R_{f_0{(M)}S}\sin^2\theta_P\nonumber\\
&+&\frac{1}{\sqrt{2}}A' R_{f_0{(M)}N'}\sin^2\theta_P
+A' R_{f_0{(M)}S'}\cos^2\theta_P+A'' R_{f_0{(M)}G}).
\end{eqnarray}
Using these coupling constants, decay widths for $f_0(M)$ are expressed as 
\begin{eqnarray}
\Gamma(f_0(M)\to \pi(m_1)+\pi(m_2))&=&\frac32\frac{\gamma^2_{f_0(M)\pi\pi}}
{32\pi}\frac{q_{Mm_1m_2}}{m_{f_0(M)}^2} m_{Mm_1m_2}^4,\nonumber\\
\Gamma(f_0(M)\to K(m_1)+\overline{K}(m_2))&=&2\frac{\gamma^2_{f_0(M)K\bar{K}}}
{32\pi}\frac{q_{Mm_1m_2}}{m_{f_0(M)}^2} m_{Mm_1m_2}^4,\nonumber\\
\Gamma(f_0(M)\to \eta(m_1)+\eta(m_2))&=&2\frac{\gamma^2_{f_0(M)\eta\eta}}
{32\pi}\frac{q_{Mm_1m_2}}{m_{f_0(M)}^2} m_{Mm_1m_2}^4,\nonumber\\
\Gamma(f_0(M)\to \eta(m_1)+\eta'(m_2))&=&\frac{\gamma^2_{f_0(M)\eta\eta'}}
{32\pi}\frac{q_{Mm_1m_2}}{m_{f_0(M)}^2} m_{Mm_1m_2}^4,\nonumber\\
\Gamma(f_0(M)\to \eta'(m_1)+\eta'(m_2))&=&2\frac{\gamma^2_{f_0(M)\eta'\eta'}}
{32\pi}\frac{q_{Mm_1m_2}}{m_{f_0(M)}^2} m_{Mm_1m_2}^4.
\end{eqnarray} 
\par
Experimental data for these decay widths and decay ratios are quoted from PDG 
\cite{pdg} and listed in second column of Table II. Using these data, we 
estimate the allowed values for $A$, $A'$ and $A''$ in the $\chi^2\le12.340$ 
corresponding to the 50\% C.L. on degree of freedom 13. 
\begin{table}
\caption{Experimental data \cite{pdg} and best fit values for various decay 
widths and decay ratios. Best fit values are obtained on $A=-2.88,\ A'=-2.28,\ 
A''=0.305,\ \theta_P=18.9^\circ$ for the $f_0(1500)$ glueball case and 
$A=-4.06,\ A'=-1.93,\ A''=0.604,\ \theta_P=50^\circ$ for the $f_0(1710)$ 
glueball case. On the $\chi^2$ fit, we did not use the data of $\Gamma_{
f_0(1370)\to {\rm all}}$, $\Gamma_{f_0(1500)\to {\rm all}}$ and $(\gamma_{
f_0(980)K\overline{K}}/\gamma_{f_0(980)\pi\pi})^2$.}  
\begin{center}
\begin{tabular}{|c|c|c|c|}\hline\hline
Decay width and ratio&Experimental data&Best fit value&
Best fit value\\ 
&&($f_0(1500)$:glueball)&($f_0(1710)$:glueball)\\ \hline
$\Gamma_{f_0(980)\to {\rm all}(\pi\pi+K\bar{K})}$&$70\pm30{\rm MeV}$&$
48{\rm MeV}$&$66{\rm MeV}$\\ \hline 
$\Gamma_{f_0(980)\to \pi\pi}/\Gamma_{f_0(980)\to {\rm all}(\pi\pi+K\bar{K})}$&
$0.74\pm0.07^{(*)}$&0.77&0.74\\ \hline 
$(\gamma_{f_0(980)K\overline{K}}/\gamma_{f_0(980)\pi\pi})^2$&$1\sim8$ 
\cite{achasov2} &5.8&6.9  \\ \hline 
$\Gamma_{f_0(1370)\to {\rm all}}$&$350\pm150{\rm MeV}$&$\Gamma_{f_0(1370)\to 
\pi\pi+K\bar{K}+\eta\eta}$&$79{\rm MeV}$\\ 
&&=165{\rm MeV}&\\ \hline 
$\Gamma_{f_0(1370)\to \pi\pi}/\Gamma_{f_0(1370)\to {\rm all}}$&
$0.26\pm0.09^{(*)}$&0.24&0.20\\ \hline
$\Gamma_{f_0(1370)\to K\bar{K}}/\Gamma_{f_0(1370)\to {\rm all}}$&
$0.35\pm0.13^{(*)}$&0.02&0.02\\ \hline
$\Gamma_{f_0(1500)\to {\rm all}}$&$109\pm7{\rm MeV}$&$\Gamma_{f_0(1500)\to 
\pi\pi+K\bar{K}+\eta\eta+\eta\eta'}$&$94{\rm MeV}$\\
&&=22{\rm MeV}&\\ \hline 
$\Gamma_{f_0(1500)\to K\bar{K}}/\Gamma_{f_0(1500)\to \pi\pi}$&
$0.19\pm0.07$&0.15&0.09\\ \hline
$\Gamma_{f_0(1500)\to \eta\eta}/\Gamma_{f_0(1500)\to \pi\pi}$&
$0.18\pm0.03$&0.17&0.22\\ \hline
$\Gamma_{f_0(1500)\to \eta\eta'}/\Gamma_{f_0(1500)\to \pi\pi}$&
$0.095\pm0.026$&0.075&0.082\\ \hline
$\Gamma_{f_0(1500)\to \eta\eta'}/\Gamma_{f_0(1500)\to \eta\eta}$&
$0.29\pm0.16$&0.43&0.38\\ \hline
$\Gamma_{f_0(1710)\to {\rm all}}$&$125\pm10{\rm MeV}$&126{\rm
 MeV}&123{\rm MeV}\\ \hline 
$\Gamma_{f_0(1710)\to \pi\pi}/\Gamma_{f_0(1710)\to K\bar{K}}$&
$0.39\pm0.14$&0.48&0.38\\ \hline
$\Gamma_{f_0(1710)\to K\bar{K}}/\Gamma_{f_0(1710)\to {\rm all}}$&
$0.38\pm0.14^{(*)}$&0.51&0.58\\ \hline
$\Gamma_{f_0(1710)\to \eta\eta}/\Gamma_{f_0(1710)\to {\rm all}}$&
$0.18\pm0.08^{(*)}$&0.17&0.19\\ \hline
$\Gamma_{f_0(1710)\to \eta\eta}/\Gamma_{f_0(1710)\to K\bar{K}}$&
$0.48\pm0.15$&0.33&0.33\\ \hline
\end{tabular}
\end{center}
\end{table}
The values with $^{(*)}$ are ones which are not decided in PDG \cite{pdg} 
and then are averaged over data cited in PDG \cite{pdg}.
We get the allowed values for $A$, $A'$ and $A''$ in the 
$\chi^2\le12.340$ for the two cases in which $f_0(1500)$ is assumed as glueball and 
$f_0(1710)$ is assumed as glueball. We did not use the data of 
$\Gamma_{f_0(1370)\to {\rm all}}$ and $\Gamma_{f_0(1500)\to {\rm all}}$ for 
$\chi^2$ fit, because the $\Gamma_{f_0(1370)\to {\rm all}}$ and $\Gamma_
{f_0(1500)\to {\rm all}}$ contain the $4\pi$ and $\rho\rho$ decays widths which 
are not included in our estimation. The allowed values for $A$, $A'$, $A''$ and 
$\theta_P$ corresponding to the $f_0(1500)$ glueball case are as follows:
\begin{equation}
\begin{array}{l}
A=-2.88\pm0.16,\ \ \ A'=-2.28\pm0.08,\ 
A''=0.305\pm0.034,\ \ \theta_P=(18.9\pm1.8)^\circ\ {\rm or}\ 
(38.8\pm0.4)^\circ,\\\end{array}\end{equation}and those corresponding to the 
$f_0(1710)$ glueball case are as follows:
\begin{equation}
\begin{array}{l}
A=-4.06\pm0.14,\ \ \ A'=-1.93\pm0.10,\ 
A''=0.640\pm0.04,\ \ \theta_P=(50\pm2)^\circ.\\
\end{array}
\end{equation}
We showed the best fit values for decay widths and decay ratios on $A=-2.88,\ 
A'=-2.28,\ A''=0.305,\ \theta_P=18.9^\circ$ for the $f_0(1500)$ glueball 
case and on $A=-4.06,\ A'=-1.93,\ A''=0.640,\ \theta_P=50^\circ$ for the 
$f_0(1710)$ glueball case in third and forth column of Table II. We listed the 
ratio $(\gamma_{f_0(980)K\overline{K}}/\gamma_{f_0(980)\pi\pi})^2$ in Table II, 
the experimental data for which is quoted from the Ref. \cite{achasov2} and is 
not used to $\chi^2$ fit. 
\par 
The characteristic features of results obtained are \\
(1) there is a large discrepancy between the value of $A$ for $I=1, 1/2$ and 
that for $I=0$,\\
(2) the value for $A'$ is $2 \sim 3$,\\
(3) the value for $A''$ is $0.3 \sim 0.6$,\\
(4) $\theta_P$ is $40^\circ \sim 50^\circ$,\\
(5) estimated values for the ratio $\Gamma_{f_0(1370)\to K\overline{K}}/
    \Gamma_{f_0(1370)\to {\rm all}}$ are small about 1 order compared to the data 
    for both $f_0(1500)$ glueball and $f_0(1710)$ glueball cases,\\
(6) estimated value for $\Gamma_{f_0(1370)\to \pi\pi+K\overline{K}+\eta\eta}$ 
    in $f_0(1710)$ glueball case seems rather small compared to the experimental 
    value for $\Gamma_{f_0(1370)\to {\rm all}}$, although experimental value 
    for $\Gamma_{f_0(1370)\to {\rm all}}$ contains $4\pi$ and $\rho\rho$ decay 
    channel and has large uncertainty.
\subsection{A resolution for discrepancy between $A$ for $I=1, 1/2$ 
and $A$ for $I=0$}
\begin{center}
\begin{figure}[htbp]
\includegraphics{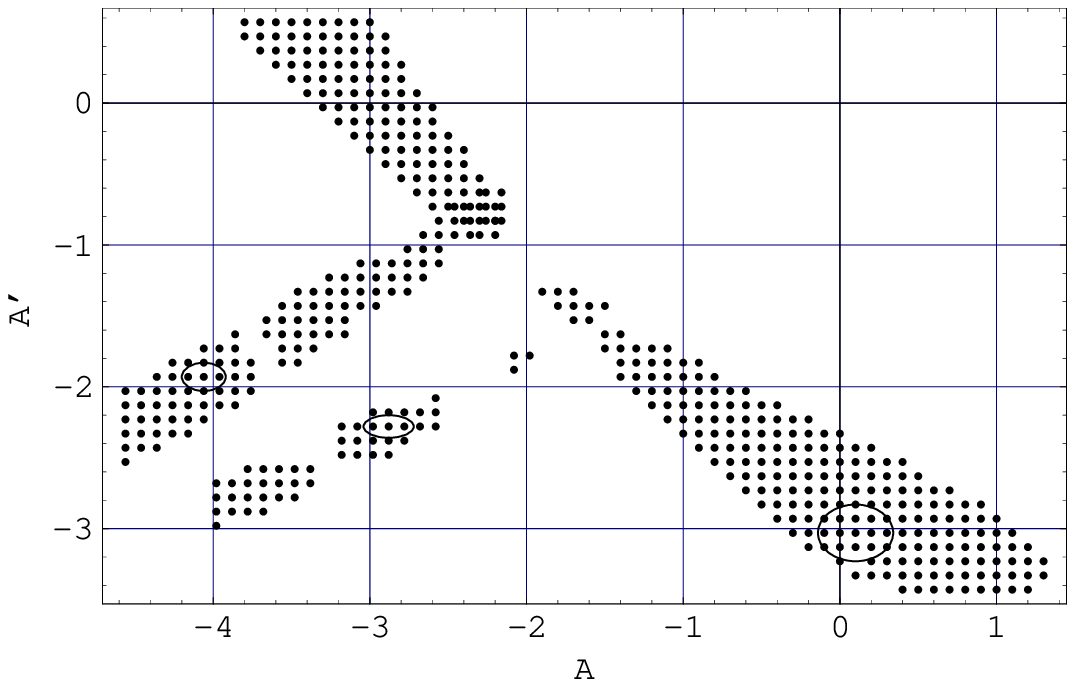}\\
Fig. 6\\
\end{figure} 
\end{center}
\par
In this subsection, we will study about a resolution for discrepancy between $A$ 
for $I=1, 1/2$ and for $I=0$. Allowed values of $(A,\ A')$ are $(0.1\pm0.24,\ 
-3.03\pm0.2)$ for $I=1, 1/2$ meson decays in 50\% C.L., and allowed values of 
$(A,\ A')$ are $(-2.88\pm0.16,\ -2.28\pm0.08)$ or $(-4.06\pm0.14,\ 
-1.93\pm0.10)$ for $I=0$ meson decays in 50\% C.L. These allowed regions are 
marked by ellipses in Fig. 6. These allowed regions are spread out 
with the increases of maximum $\chi^2$  from 5.348 to 20 in $\chi^2$ fit of 
$I=1,\ 1/2$ meson decays and from 12.340 to 30 in $\chi^2$ fit of $I=0$ meson 
decays as shown in fig.~6. The allowed region for $I=1,\ 1/2$ meson decays is 
spread out to up-left direction from down-right region $(0.1\pm0.24,\ 
-3.03\pm0.2)$ in $(A,\ A')$ plane. The allowed region for $I=0$ meson decays 
$f_0(1500)$ glueball case  is spread out to center direction from down-left 
region $(-2.88\pm0.16,\ -2.28\pm0.08)$ in $(A,\ A')$ plane, and the allowed 
region for $I=0$ meson decays $f_0(1710)$ glueball case  is spread out to 
center direction from down-left region $(-4.06\pm0.14,\ -1.93\pm0.10)$ in 
$(A,\ A')$ plane. The spread region for $I=1, 1/2$ meson decay joins with 
the spread region for $I=0$ meson decay $f_0(1710)$ glueball case at $(\sim
-2.3, \sim-0.8)$ in $(A,\ A')$ plane. Also, the spread region for $I=1, 1/2$ 
meson decay can join with the spread region for $I=0$ meson decay $f_0(1500)$ 
glueball at $(\sim-1.7,\ \sim-1.4)$ in $(A,\ A')$ plane if maximum $\chi^2$ 
increases moreover in $\chi^2$ fit.
\section{Conclusion}
From the analysis  of the mass spectroscopy of low mass and high mass scalar 
mesons, we can get the result that the coupling strengths of the mixing between 
low mass and high mass scalar mesons are very strong and strengths of the 
coupling for $I=1$, $L=1/2$ and $I=0$ are almost same. We further analyze the 
glueball mixing among these scalar mesons. We get the mixing parameters of 
$I=0$ high mass $f_0(1370)$, $f_0(1500)$ and $f_0(1710)$ mesons, in which 
$f_0(1500)$ is considered to be the glueball or $f_0(1710)$ is considered to
 be the glueball. 
\par 
The strong mixing between the low and high mass scalar mesons affects the decay 
processes of these scalar mesons. From the analysis of the decay widths and 
decay ratios of the low and high mass scalar mesons, we got the results that 
the coupling constants $A$ for $N\to PP$ and $A'$ for $N'\to PP$ 
are about $(0.10\pm0.24,\ -3.03\pm0.2)$ for $I=1,\ 1/2$ mesons on 50\% C.L. 
The results obtained for $I=0$ mesons are $(A,\ A',\ A'')=(-2.88\pm0.16,\ 
-2.28\pm0.08,\ 0.305\pm0.034)$ or $(-4.06\pm0.14,\ -1.93\pm0.10,\ 0.640\pm
0.04)$ on 50\% C.L., for $f_0(1500)$ glueball case or $f_0(1710)$ glueball 
case, respectively. Here, $A''$ is the coupling constant for $G\to PP$. 
The $\theta_P$ is obtained to be $40^\circ \sim 50^\circ$. The large discrepancy 
between the allowed values $A$ as $\sim 0.10$ for $I=1,\ 1/2$ and $\sim-2.88$ or 
$\sim-4.06$ for $I=0$ can be resolved if one increases the maximum $\chi^2$ 
values in $\chi^2$ fit estimating the allowed regions and gets the extended 
allowed regions. The extension of allowed regions gives the common allowed 
values $(A,\ A')\sim(-2.3,\ -0.8)$. \\

\end{document}

%% file: fig1.tex
\unitlength 0.1in
\begin{picture}( 25.0000,  6.1000)(  2.0000,-10.1000)
%
\special{pn 13}%
\special{pa 600 410}%
\special{pa 2600 410}%
\special{fp}%
\special{pa 2600 1010}%
\special{pa 600 1010}%
\special{fp}%
\special{pa 600 610}%
\special{pa 1300 610}%
\special{fp}%
\special{pa 1300 810}%
\special{pa 600 810}%
\special{fp}%
\special{pa 2600 610}%
\special{pa 1900 610}%
\special{fp}%
\special{pa 1900 810}%
\special{pa 2600 810}%
\special{fp}%
%
\special{pn 13}%
\special{ar 1900 710 100 100  1.5707963 4.7123890}%
%
\special{pn 13}%
\special{ar 1300 710 100 100  4.7123890 6.2831853}%
\special{ar 1300 710 100 100  0.0000000 1.5707963}%
\special{pn 13}%
\special{pa 1400 700}%
\special{pa 1406 704}%
\special{pa 1410 708}%
\special{pa 1416 714}%
\special{pa 1420 718}%
\special{pa 1426 722}%
\special{pa 1430 726}%
\special{pa 1436 728}%
\special{pa 1440 730}%
\special{pa 1446 730}%
\special{pa 1450 730}%
\special{pa 1456 726}%
\special{pa 1460 724}%
\special{pa 1466 720}%
\special{pa 1470 714}%
\special{pa 1476 710}%
\special{pa 1480 704}%
\special{pa 1486 700}%
\special{pa 1490 696}%
\special{pa 1496 694}%
\special{pa 1500 692}%
\special{pa 1506 690}%
\special{pa 1510 690}%
\special{pa 1516 692}%
\special{pa 1520 696}%
\special{pa 1526 700}%
\special{pa 1530 704}%
\special{pa 1536 708}%
\special{pa 1540 714}%
\special{pa 1546 718}%
\special{pa 1550 722}%
\special{pa 1556 726}%
\special{pa 1560 728}%
\special{pa 1566 730}%
\special{pa 1570 730}%
\special{pa 1576 730}%
\special{pa 1580 728}%
\special{pa 1586 724}%
\special{pa 1590 720}%
\special{pa 1596 716}%
\special{pa 1600 710}%
\special{pa 1606 706}%
\special{pa 1610 700}%
\special{pa 1616 696}%
\special{pa 1620 694}%
\special{pa 1626 692}%
\special{pa 1630 690}%
\special{pa 1636 690}%
\special{pa 1640 692}%
\special{pa 1646 694}%
\special{pa 1650 698}%
\special{pa 1656 702}%
\special{pa 1660 708}%
\special{pa 1666 712}%
\special{pa 1670 718}%
\special{pa 1676 722}%
\special{pa 1680 726}%
\special{pa 1686 728}%
\special{pa 1690 730}%
\special{pa 1696 730}%
\special{pa 1700 730}%
\special{pa 1706 728}%
\special{pa 1710 724}%
\special{pa 1716 720}%
\special{pa 1720 716}%
\special{pa 1726 712}%
\special{pa 1730 706}%
\special{pa 1736 702}%
\special{pa 1740 698}%
\special{pa 1746 694}%
\special{pa 1750 692}%
\special{pa 1756 690}%
\special{pa 1760 690}%
\special{pa 1766 692}%
\special{pa 1770 694}%
\special{pa 1776 698}%
\special{pa 1780 702}%
\special{pa 1786 708}%
\special{pa 1790 712}%
\special{pa 1796 716}%
\special{pa 1800 722}%
\special{sp}%
%
\special{pn 13}%
\special{pa 600 410}%
\special{pa 1200 410}%
\special{pa 1200 1010}%
\special{pa 600 1010}%
\special{pa 600 410}%
\special{ip}%
%
\special{pn 13}%
\special{pa 590 410}%
\special{pa 1190 410}%
\special{pa 1190 1010}%
\special{pa 590 1010}%
\special{pa 590 410}%
\special{ip}%
%
\special{pn 4}%
\special{pa 1120 810}%
\special{pa 930 1000}%
\special{fp}%
\special{pa 1000 810}%
\special{pa 810 1000}%
\special{fp}%
\special{pa 880 810}%
\special{pa 690 1000}%
\special{fp}%
\special{pa 760 810}%
\special{pa 600 970}%
\special{fp}%
\special{pa 640 810}%
\special{pa 600 850}%
\special{fp}%
\special{pa 1180 870}%
\special{pa 1050 1000}%
\special{fp}%
%
\special{pn 4}%
\special{pa 1080 610}%
\special{pa 890 800}%
\special{fp}%
\special{pa 960 610}%
\special{pa 770 800}%
\special{fp}%
\special{pa 840 610}%
\special{pa 650 800}%
\special{fp}%
\special{pa 720 610}%
\special{pa 600 730}%
\special{fp}%
\special{pa 1180 630}%
\special{pa 1010 800}%
\special{fp}%
\special{pa 1180 750}%
\special{pa 1130 800}%
\special{fp}%
%
\special{pn 4}%
\special{pa 1040 410}%
\special{pa 850 600}%
\special{fp}%
\special{pa 920 410}%
\special{pa 730 600}%
\special{fp}%
\special{pa 800 410}%
\special{pa 610 600}%
\special{fp}%
\special{pa 680 410}%
\special{pa 600 490}%
\special{fp}%
\special{pa 1160 410}%
\special{pa 970 600}%
\special{fp}%
\special{pa 1180 510}%
\special{pa 1090 600}%
\special{fp}%
%
\special{pn 8}%
\special{pa 1200 410}%
\special{pa 1200 1010}%
\special{fp}%
\special{pa 2000 1010}%
\special{pa 2000 410}%
\special{fp}%
%
\special{pn 8}%
\special{pa 2000 410}%
\special{pa 2600 410}%
\special{pa 2600 1010}%
\special{pa 2000 1010}%
\special{pa 2000 410}%
\special{ip}%
%
\special{pn 8}%
\special{pa 2000 410}%
\special{pa 2600 410}%
\special{pa 2600 1010}%
\special{pa 2000 1010}%
\special{pa 2000 410}%
\special{ip}%
%
\special{pn 4}%
\special{pa 2440 810}%
\special{pa 2250 1000}%
\special{fp}%
\special{pa 2320 810}%
\special{pa 2130 1000}%
\special{fp}%
\special{pa 2200 810}%
\special{pa 2010 1000}%
\special{fp}%
\special{pa 2080 810}%
\special{pa 2000 890}%
\special{fp}%
\special{pa 2560 810}%
\special{pa 2370 1000}%
\special{fp}%
\special{pa 2600 890}%
\special{pa 2490 1000}%
\special{fp}%
%
\special{pn 8}%
\special{pa 2000 400}%
\special{pa 2600 400}%
\special{pa 2600 800}%
\special{pa 2000 800}%
\special{pa 2000 400}%
\special{ip}%
%
\special{pn 4}%
\special{pa 2280 610}%
\special{pa 2090 800}%
\special{fp}%
\special{pa 2160 610}%
\special{pa 2000 770}%
\special{fp}%
\special{pa 2040 610}%
\special{pa 2000 650}%
\special{fp}%
\special{pa 2400 610}%
\special{pa 2210 800}%
\special{fp}%
\special{pa 2520 610}%
\special{pa 2330 800}%
\special{fp}%
\special{pa 2600 650}%
\special{pa 2450 800}%
\special{fp}%
%
\special{pn 4}%
\special{pa 2360 410}%
\special{pa 2170 600}%
\special{fp}%
\special{pa 2240 410}%
\special{pa 2050 600}%
\special{fp}%
\special{pa 2120 410}%
\special{pa 2000 530}%
\special{fp}%
\special{pa 2480 410}%
\special{pa 2290 600}%
\special{fp}%
\special{pa 2590 420}%
\special{pa 2410 600}%
\special{fp}%
\special{pa 2600 530}%
\special{pa 2530 600}%
\special{fp}%
%
\special{pn 13}%
\special{pa 1200 610}%
\special{pa 1300 610}%
\special{fp}%
\special{sh 1}%
\special{pa 1300 610}%
\special{pa 1234 590}%
\special{pa 1248 610}%
\special{pa 1234 630}%
\special{pa 1300 610}%
\special{fp}%
\special{pa 1300 810}%
\special{pa 1200 810}%
\special{fp}%
\special{sh 1}%
\special{pa 1200 810}%
\special{pa 1268 830}%
\special{pa 1254 810}%
\special{pa 1268 790}%
\special{pa 1200 810}%
\special{fp}%
\special{pa 1880 610}%
\special{pa 1980 610}%
\special{fp}%
\special{sh 1}%
\special{pa 1980 610}%
\special{pa 1914 590}%
\special{pa 1928 610}%
\special{pa 1914 630}%
\special{pa 1980 610}%
\special{fp}%
\special{pa 1980 810}%
\special{pa 1890 810}%
\special{fp}%
\special{sh 1}%
\special{pa 1890 810}%
\special{pa 1958 830}%
\special{pa 1944 810}%
\special{pa 1958 790}%
\special{pa 1890 810}%
\special{fp}%
\put(27.0000,-7.6000){\makebox(0,0)[lb]{$qq\bar{q}\bar{q}$}}%
\put(2.0000,-7.6000){\makebox(0,0)[lb]{$qq\bar{q}\bar{q}$}}%
%
\special{pn 13}%
\special{pa 1800 1010}%
\special{pa 1600 1010}%
\special{fp}%
\special{sh 1}%
\special{pa 1600 1010}%
\special{pa 1668 1030}%
\special{pa 1654 1010}%
\special{pa 1668 990}%
\special{pa 1600 1010}%
\special{fp}%
\special{pa 1400 410}%
\special{pa 1600 410}%
\special{fp}%
\special{sh 1}%
\special{pa 1600 410}%
\special{pa 1534 390}%
\special{pa 1548 410}%
\special{pa 1534 430}%
\special{pa 1600 410}%
\special{fp}%
\end{picture}%

%% file: fig2.tex
\unitlength 0.1in
\begin{picture}( 24.5000,  6.1000)(  2.4000, -8.0000)
%
\special{pn 13}%
\special{pa 600 200}%
\special{pa 2600 200}%
\special{fp}%
\special{pa 2600 800}%
\special{pa 600 800}%
\special{fp}%
\special{pa 600 600}%
\special{pa 1500 600}%
\special{fp}%
\special{pa 1500 400}%
\special{pa 600 400}%
\special{fp}%
%
\special{pn 13}%
\special{ar 1500 500 100 100  4.7123890 6.2831853}%
\special{ar 1500 500 100 100  0.0000000 1.5707963}%
%
\special{pn 8}%
\special{pa 1300 200}%
\special{pa 1300 800}%
\special{fp}%
\special{pa 1900 800}%
\special{pa 1900 200}%
\special{fp}%
%
\special{pn 8}%
\special{pa 1900 200}%
\special{pa 2610 200}%
\special{pa 2610 800}%
\special{pa 1900 800}%
\special{pa 1900 200}%
\special{ip}%
%
\special{pn 4}%
\special{pa 2600 200}%
\special{pa 2010 790}%
\special{fp}%
\special{pa 2480 200}%
\special{pa 1900 780}%
\special{fp}%
\special{pa 2360 200}%
\special{pa 1900 660}%
\special{fp}%
\special{pa 2240 200}%
\special{pa 1900 540}%
\special{fp}%
\special{pa 2120 200}%
\special{pa 1900 420}%
\special{fp}%
\special{pa 2000 200}%
\special{pa 1900 300}%
\special{fp}%
\special{pa 2610 310}%
\special{pa 2130 790}%
\special{fp}%
\special{pa 2610 430}%
\special{pa 2250 790}%
\special{fp}%
\special{pa 2610 550}%
\special{pa 2370 790}%
\special{fp}%
\special{pa 2610 670}%
\special{pa 2490 790}%
\special{fp}%
%
\special{pn 8}%
\special{pa 590 190}%
\special{pa 1300 190}%
\special{pa 1300 790}%
\special{pa 590 790}%
\special{pa 590 190}%
\special{ip}%
%
\special{pn 4}%
\special{pa 1120 600}%
\special{pa 930 790}%
\special{fp}%
\special{pa 1000 600}%
\special{pa 810 790}%
\special{fp}%
\special{pa 880 600}%
\special{pa 690 790}%
\special{fp}%
\special{pa 760 600}%
\special{pa 590 770}%
\special{fp}%
\special{pa 640 600}%
\special{pa 590 650}%
\special{fp}%
\special{pa 1240 600}%
\special{pa 1050 790}%
\special{fp}%
\special{pa 1300 660}%
\special{pa 1170 790}%
\special{fp}%
%
\special{pn 4}%
\special{pa 1080 400}%
\special{pa 890 590}%
\special{fp}%
\special{pa 960 400}%
\special{pa 770 590}%
\special{fp}%
\special{pa 840 400}%
\special{pa 650 590}%
\special{fp}%
\special{pa 720 400}%
\special{pa 590 530}%
\special{fp}%
\special{pa 1200 400}%
\special{pa 1010 590}%
\special{fp}%
\special{pa 1300 420}%
\special{pa 1130 590}%
\special{fp}%
\special{pa 1300 540}%
\special{pa 1250 590}%
\special{fp}%
%
\special{pn 4}%
\special{pa 1040 200}%
\special{pa 850 390}%
\special{fp}%
\special{pa 920 200}%
\special{pa 730 390}%
\special{fp}%
\special{pa 800 200}%
\special{pa 610 390}%
\special{fp}%
\special{pa 680 200}%
\special{pa 590 290}%
\special{fp}%
\special{pa 1160 200}%
\special{pa 970 390}%
\special{fp}%
\special{pa 1280 200}%
\special{pa 1090 390}%
\special{fp}%
\special{pa 1300 300}%
\special{pa 1210 390}%
\special{fp}%
\special{pn 8}%
\special{pn 13}%
\special{pa 1300 514}%
\special{pa 1306 516}%
\special{pa 1310 520}%
\special{pa 1316 520}%
\special{pa 1320 520}%
\special{pa 1326 520}%
\special{pa 1330 516}%
\special{pa 1336 514}%
\special{pa 1340 508}%
\special{pa 1346 504}%
\special{pa 1350 500}%
\special{pa 1356 494}%
\special{pa 1360 490}%
\special{pa 1366 486}%
\special{pa 1370 482}%
\special{pa 1376 482}%
\special{pa 1380 480}%
\special{pa 1386 482}%
\special{pa 1390 482}%
\special{pa 1396 486}%
\special{pa 1400 490}%
\special{pa 1406 494}%
\special{pa 1410 498}%
\special{pa 1416 504}%
\special{pa 1420 508}%
\special{pa 1426 512}%
\special{pa 1430 516}%
\special{pa 1436 518}%
\special{pa 1440 520}%
\special{pa 1446 520}%
\special{pa 1450 520}%
\special{pa 1456 516}%
\special{pa 1460 514}%
\special{pa 1466 510}%
\special{pa 1470 504}%
\special{pa 1476 500}%
\special{pa 1480 494}%
\special{pa 1486 490}%
\special{pa 1490 486}%
\special{pa 1496 484}%
\special{pa 1500 482}%
\special{pa 1506 480}%
\special{pa 1510 480}%
\special{pa 1516 482}%
\special{pa 1520 486}%
\special{pa 1526 490}%
\special{pa 1530 494}%
\special{pa 1536 498}%
\special{pa 1540 504}%
\special{pa 1546 508}%
\special{pa 1550 512}%
\special{pa 1556 516}%
\special{pa 1560 518}%
\special{pa 1566 520}%
\special{pa 1570 520}%
\special{pa 1576 520}%
\special{pa 1580 518}%
\special{pa 1586 514}%
\special{pa 1590 510}%
\special{pa 1596 506}%
\special{ip}%
\special{pa 1600 500}%
\special{pa 1606 496}%
\special{pa 1610 490}%
\special{pa 1616 486}%
\special{pa 1620 484}%
\special{pa 1626 482}%
\special{pa 1630 480}%
\special{pa 1636 480}%
\special{pa 1640 482}%
\special{pa 1646 484}%
\special{pa 1650 488}%
\special{pa 1656 492}%
\special{pa 1660 498}%
\special{pa 1666 502}%
\special{pa 1670 508}%
\special{pa 1676 512}%
\special{pa 1680 516}%
\special{pa 1686 518}%
\special{pa 1690 520}%
\special{pa 1696 520}%
\special{pa 1700 520}%
\special{pa 1706 518}%
\special{pa 1710 514}%
\special{pa 1716 510}%
\special{pa 1720 506}%
\special{pa 1726 502}%
\special{pa 1730 496}%
\special{pa 1736 492}%
\special{pa 1740 488}%
\special{pa 1746 484}%
\special{pa 1750 482}%
\special{pa 1756 480}%
\special{pa 1760 480}%
\special{pa 1766 482}%
\special{pa 1770 484}%
\special{pa 1776 488}%
\special{pa 1780 492}%
\special{pa 1786 498}%
\special{pa 1790 502}%
\special{pa 1796 506}%
\special{pa 1800 512}%
\special{pa 1806 516}%
\special{pa 1810 518}%
\special{pa 1816 520}%
\special{pa 1820 520}%
\special{pa 1826 520}%
\special{pa 1830 518}%
\special{pa 1836 516}%
\special{pa 1840 512}%
\special{pa 1846 506}%
\special{pa 1850 502}%
\special{pa 1856 496}%
\special{pa 1860 492}%
\special{pa 1866 488}%
\special{pa 1870 484}%
\special{pa 1876 482}%
\special{pa 1880 480}%
\special{pa 1886 480}%
\special{pa 1890 482}%
\special{pa 1896 484}%
\special{pa 1900 488}%
\special{sp}%
%
\special{pn 13}%
\special{pa 1400 200}%
\special{pa 1600 200}%
\special{fp}%
\special{sh 1}%
\special{pa 1600 200}%
\special{pa 1534 180}%
\special{pa 1548 200}%
\special{pa 1534 220}%
\special{pa 1600 200}%
\special{fp}%
\special{pa 1800 800}%
\special{pa 1600 800}%
\special{fp}%
\special{sh 1}%
\special{pa 1600 800}%
\special{pa 1668 820}%
\special{pa 1654 800}%
\special{pa 1668 780}%
\special{pa 1600 800}%
\special{fp}%
\special{pa 1350 600}%
\special{pa 1450 600}%
\special{fp}%
\special{sh 1}%
\special{pa 1450 600}%
\special{pa 1384 580}%
\special{pa 1398 600}%
\special{pa 1384 620}%
\special{pa 1450 600}%
\special{fp}%
\special{pa 1460 400}%
\special{pa 1370 400}%
\special{fp}%
\special{sh 1}%
\special{pa 1370 400}%
\special{pa 1438 420}%
\special{pa 1424 400}%
\special{pa 1438 380}%
\special{pa 1370 400}%
\special{fp}%
\put(2.4000,-5.4000){\makebox(0,0)[lb]{$qq\bar{q}\bar{q}$}}%
\put(26.9000,-5.4000){\makebox(0,0)[lb]{$q\bar{q}$}}%
\end{picture}%

%% file: fig3.tex
\unitlength 0.1in
\begin{picture}( 26.0000, 11.0300)(  3.0000,-12.3300)
\special{pn 13}%
\special{pa 1416 414}%
\special{pa 1420 410}%
\special{pa 1426 406}%
\special{pa 1430 400}%
\special{pa 1436 396}%
\special{pa 1440 394}%
\special{pa 1446 390}%
\special{pa 1450 388}%
\special{pa 1456 388}%
\special{pa 1460 386}%
\special{pa 1466 386}%
\special{pa 1470 388}%
\special{pa 1476 390}%
\special{pa 1480 392}%
\special{pa 1486 396}%
\special{pa 1490 400}%
\special{pa 1496 404}%
\special{pa 1500 408}%
\special{pa 1506 412}%
\special{pa 1510 416}%
\special{pa 1516 422}%
\special{pa 1520 424}%
\special{pa 1526 428}%
\special{pa 1530 430}%
\special{pa 1536 432}%
\special{pa 1540 434}%
\special{pa 1546 434}%
\special{pa 1550 434}%
\special{pa 1556 432}%
\special{pa 1560 430}%
\special{pa 1566 428}%
\special{pa 1570 424}%
\special{pa 1576 420}%
\special{pa 1580 416}%
\special{pa 1586 412}%
\special{pa 1590 408}%
\special{pa 1596 404}%
\special{pa 1600 400}%
\special{pa 1606 396}%
\special{pa 1610 392}%
\special{pa 1616 390}%
\special{pa 1620 388}%
\special{pa 1626 386}%
\special{pa 1630 386}%
\special{pa 1636 388}%
\special{pa 1640 388}%
\special{pa 1646 390}%
\special{pa 1650 394}%
\special{pa 1656 398}%
\special{pa 1660 402}%
\special{pa 1666 406}%
\special{pa 1670 410}%
\special{pa 1676 414}%
\special{pa 1680 418}%
\special{pa 1686 422}%
\special{pa 1690 426}%
\special{pa 1696 430}%
\special{pa 1700 432}%
\special{pa 1706 434}%
\special{pa 1710 434}%
\special{pa 1716 434}%
\special{pa 1720 434}%
\special{pa 1726 432}%
\special{pa 1730 430}%
\special{pa 1736 426}%
\special{pa 1740 422}%
\special{pa 1746 418}%
\special{pa 1750 414}%
\special{pa 1756 410}%
\special{pa 1760 406}%
\special{pa 1766 402}%
\special{pa 1770 398}%
\special{pa 1776 394}%
\special{pa 1780 390}%
\special{pa 1786 388}%
\special{pa 1790 388}%
\special{pa 1796 386}%
\special{pa 1800 386}%
\special{pa 1806 388}%
\special{pa 1810 390}%
\special{pa 1816 392}%
\special{pa 1820 396}%
\special{pa 1826 398}%
\special{pa 1830 404}%
\special{pa 1836 408}%
\special{pa 1840 412}%
\special{pa 1846 416}%
\special{pa 1850 420}%
\special{pa 1856 424}%
\special{pa 1860 428}%
\special{pa 1866 430}%
\special{pa 1870 432}%
\special{pa 1876 434}%
\special{pa 1880 434}%
\special{pa 1886 434}%
\special{pa 1890 432}%
\special{pa 1896 430}%
\special{pa 1900 428}%
\special{pa 1906 424}%
\special{pa 1910 422}%
\special{pa 1916 416}%
\special{pa 1920 412}%
\special{pa 1926 408}%
\special{pa 1930 404}%
\special{pa 1936 400}%
\special{pa 1940 396}%
\special{pa 1946 392}%
\special{pa 1950 390}%
\special{pa 1956 388}%
\special{sp}%
%
\special{pn 13}%
\special{pa 620 396}%
\special{pa 1426 396}%
\special{fp}%
\special{pa 1426 996}%
\special{pa 1426 996}%
\special{fp}%
\special{pa 620 996}%
\special{pa 620 996}%
\special{fp}%
\special{pa 620 996}%
\special{pa 1412 996}%
\special{fp}%
\special{pa 1412 996}%
\special{pa 1412 396}%
\special{fp}%
\special{pa 1948 396}%
\special{pa 2754 396}%
\special{fp}%
\special{pa 2754 996}%
\special{pa 1948 996}%
\special{fp}%
\special{pa 1948 996}%
\special{pa 1948 396}%
\special{fp}%
\special{pn 13}%
\special{pa 1406 1008}%
\special{pa 1410 1004}%
\special{pa 1416 1000}%
\special{pa 1420 996}%
\special{pa 1426 992}%
\special{pa 1430 988}%
\special{pa 1436 984}%
\special{pa 1440 980}%
\special{pa 1446 976}%
\special{pa 1450 974}%
\special{pa 1456 972}%
\special{pa 1460 970}%
\special{pa 1466 970}%
\special{pa 1470 972}%
\special{pa 1476 972}%
\special{pa 1480 974}%
\special{pa 1486 978}%
\special{pa 1490 980}%
\special{pa 1496 984}%
\special{pa 1500 990}%
\special{pa 1506 994}%
\special{pa 1510 998}%
\special{pa 1516 1002}%
\special{pa 1520 1006}%
\special{pa 1526 1010}%
\special{pa 1530 1014}%
\special{pa 1536 1016}%
\special{pa 1540 1018}%
\special{pa 1546 1018}%
\special{pa 1550 1018}%
\special{pa 1556 1018}%
\special{pa 1560 1016}%
\special{pa 1566 1014}%
\special{pa 1570 1010}%
\special{pa 1576 1008}%
\special{pa 1580 1004}%
\special{pa 1586 998}%
\special{pa 1590 994}%
\special{pa 1596 990}%
\special{pa 1600 986}%
\special{pa 1606 982}%
\special{pa 1610 978}%
\special{pa 1616 974}%
\special{pa 1620 972}%
\special{pa 1626 972}%
\special{pa 1630 970}%
\special{pa 1636 970}%
\special{pa 1640 972}%
\special{pa 1646 974}%
\special{pa 1650 976}%
\special{pa 1656 980}%
\special{pa 1660 982}%
\special{pa 1666 986}%
\special{pa 1670 992}%
\special{pa 1676 996}%
\special{pa 1680 1000}%
\special{pa 1686 1004}%
\special{pa 1690 1008}%
\special{pa 1696 1012}%
\special{pa 1700 1014}%
\special{pa 1706 1016}%
\special{pa 1710 1018}%
\special{pa 1716 1018}%
\special{pa 1720 1018}%
\special{pa 1726 1018}%
\special{pa 1730 1016}%
\special{pa 1736 1012}%
\special{pa 1740 1010}%
\special{pa 1746 1006}%
\special{pa 1750 1002}%
\special{pa 1756 996}%
\special{pa 1760 992}%
\special{pa 1766 988}%
\special{pa 1770 984}%
\special{pa 1776 980}%
\special{pa 1780 976}%
\special{pa 1786 974}%
\special{pa 1790 972}%
\special{pa 1796 970}%
\special{pa 1800 970}%
\special{pa 1806 970}%
\special{pa 1810 972}%
\special{pa 1816 974}%
\special{pa 1820 978}%
\special{pa 1826 980}%
\special{pa 1830 984}%
\special{pa 1836 988}%
\special{pa 1840 994}%
\special{pa 1846 998}%
\special{pa 1850 1002}%
\special{pa 1856 1006}%
\special{pa 1860 1010}%
\special{pa 1866 1014}%
\special{pa 1870 1016}%
\special{pa 1876 1018}%
\special{pa 1880 1018}%
\special{pa 1886 1018}%
\special{pa 1890 1018}%
\special{pa 1896 1016}%
\special{pa 1900 1014}%
\special{pa 1906 1012}%
\special{pa 1910 1008}%
\special{pa 1916 1004}%
\special{pa 1920 1000}%
\special{pa 1926 994}%
\special{pa 1930 990}%
\special{pa 1936 986}%
\special{pa 1940 982}%
\special{sp}%
%
\special{pn 8}%
\special{pa 1136 410}%
\special{pa 600 410}%
\special{pa 600 1012}%
\special{pa 1136 1012}%
\special{pa 1136 410}%
\special{ip}%
%
\special{pn 8}%
\special{pa 2224 410}%
\special{pa 2760 410}%
\special{pa 2760 1012}%
\special{pa 2224 1012}%
\special{pa 2224 410}%
\special{ip}%
%
\special{pn 8}%
\special{pa 2760 458}%
\special{pa 2224 940}%
\special{fp}%
\special{pa 2760 604}%
\special{pa 2316 998}%
\special{fp}%
\special{pa 2760 746}%
\special{pa 2478 998}%
\special{fp}%
\special{pa 2760 890}%
\special{pa 2640 998}%
\special{fp}%
\special{pa 2654 410}%
\special{pa 2224 796}%
\special{fp}%
\special{pa 2490 410}%
\special{pa 2224 650}%
\special{fp}%
\special{pa 2332 410}%
\special{pa 2224 506}%
\special{fp}%
%
\special{pn 4}%
\special{pa 1136 614}%
\special{pa 710 998}%
\special{fp}%
\special{pa 1136 472}%
\special{pa 600 950}%
\special{fp}%
\special{pa 1044 410}%
\special{pa 600 806}%
\special{fp}%
\special{pa 882 410}%
\special{pa 600 662}%
\special{fp}%
\special{pa 722 410}%
\special{pa 600 518}%
\special{fp}%
\special{pa 1136 758}%
\special{pa 868 998}%
\special{fp}%
\special{pa 1136 904}%
\special{pa 1030 998}%
\special{fp}%
%
\special{pn 8}%
\special{pa 2236 410}%
\special{pa 2236 1012}%
\special{ip}%
%
\special{pn 8}%
\special{pa 2214 394}%
\special{pa 2214 996}%
\special{fp}%
\special{pa 1140 996}%
\special{pa 1140 394}%
\special{fp}%
\put(3.0000,-8.0000){\makebox(0,0)[lb]{$q\bar{q}$}}%
\put(29.0000,-8.0000){\makebox(0,0)[lb]{$q\bar{q}$}}%
\put(16.0000,-3.0000){\makebox(0,0)[lb]{$g$}}%
\put(16.0000,-12.5000){\makebox(0,0)[lb]{$g$}}%
%
\special{pn 13}%
\special{pa 990 400}%
\special{pa 1000 400}%
\special{fp}%
\special{sh 1}%
\special{pa 1000 400}%
\special{pa 934 380}%
\special{pa 948 400}%
\special{pa 934 420}%
\special{pa 1000 400}%
\special{fp}%
%
\special{pn 13}%
\special{pa 930 1000}%
\special{pa 920 1000}%
\special{fp}%
\special{sh 1}%
\special{pa 920 1000}%
\special{pa 988 1020}%
\special{pa 974 1000}%
\special{pa 988 980}%
\special{pa 920 1000}%
\special{fp}%
%
\special{pn 13}%
\special{pa 2390 400}%
\special{pa 2400 400}%
\special{fp}%
\special{sh 1}%
\special{pa 2400 400}%
\special{pa 2334 380}%
\special{pa 2348 400}%
\special{pa 2334 420}%
\special{pa 2400 400}%
\special{fp}%
%
\special{pn 13}%
\special{pa 2330 1000}%
\special{pa 2320 1000}%
\special{fp}%
\special{sh 1}%
\special{pa 2320 1000}%
\special{pa 2388 1020}%
\special{pa 2374 1000}%
\special{pa 2388 980}%
\special{pa 2320 1000}%
\special{fp}%
\end{picture}%

%% file: fig4.tex
\unitlength 0.1in
\begin{picture}( 50.4000, 10.8000)(  3.0000,-12.5000)
%
\special{pn 13}%
\special{pa 600 410}%
\special{pa 1546 410}%
\special{fp}%
\special{pa 1546 1010}%
\special{pa 600 1010}%
\special{fp}%
\special{pa 1546 1010}%
\special{pa 1546 410}%
\special{fp}%
%
\special{pn 13}%
\special{pa 600 410}%
\special{pa 600 410}%
\special{pa 600 1010}%
\special{pa 600 1010}%
\special{pa 600 410}%
\special{ip}%
%
\special{pn 13}%
\special{pa 600 1010}%
\special{pa 1310 1010}%
\special{pa 1310 410}%
\special{pa 600 410}%
\special{pa 600 1010}%
\special{ip}%
\special{pn 8}%
\special{pa 1546 412}%
\special{pa 1550 406}%
\special{pa 1556 402}%
\special{pa 1560 398}%
\special{pa 1566 394}%
\special{pa 1570 390}%
\special{pa 1576 388}%
\special{pa 1580 386}%
\special{pa 1586 386}%
\special{pa 1590 388}%
\special{pa 1596 390}%
\special{pa 1600 392}%
\special{pa 1606 396}%
\special{pa 1610 400}%
\special{pa 1616 406}%
\special{pa 1620 410}%
\special{pa 1626 416}%
\special{pa 1630 420}%
\special{pa 1636 424}%
\special{pa 1640 428}%
\special{pa 1646 432}%
\special{pa 1650 434}%
\special{pa 1656 434}%
\special{pa 1660 434}%
\special{pa 1666 434}%
\special{pa 1670 430}%
\special{pa 1676 428}%
\special{pa 1680 424}%
\special{pa 1686 420}%
\special{pa 1690 414}%
\special{pa 1696 410}%
\special{pa 1700 404}%
\special{pa 1706 400}%
\special{pa 1710 396}%
\special{pa 1716 392}%
\special{pa 1720 388}%
\special{pa 1726 388}%
\special{pa 1730 386}%
\special{pa 1736 386}%
\special{pa 1740 388}%
\special{pa 1746 390}%
\special{pa 1750 394}%
\special{pa 1756 398}%
\special{pa 1760 402}%
\special{pa 1766 408}%
\special{pa 1770 414}%
\special{pa 1776 418}%
\special{pa 1780 422}%
\special{pa 1786 426}%
\special{pa 1790 430}%
\special{pa 1796 432}%
\special{pa 1800 434}%
\special{pa 1806 434}%
\special{pa 1810 434}%
\special{pa 1816 432}%
\special{pa 1820 430}%
\special{pa 1826 426}%
\special{pa 1830 422}%
\special{pa 1836 416}%
\special{pa 1840 412}%
\special{pa 1846 406}%
\special{pa 1850 402}%
\special{pa 1856 398}%
\special{pa 1860 394}%
\special{pa 1866 390}%
\special{pa 1870 388}%
\special{pa 1876 386}%
\special{pa 1880 386}%
\special{pa 1886 388}%
\special{pa 1890 390}%
\special{pa 1896 392}%
\special{pa 1900 396}%
\special{pa 1906 400}%
\special{pa 1910 406}%
\special{pa 1916 410}%
\special{pa 1920 416}%
\special{pa 1926 420}%
\special{pa 1930 424}%
\special{pa 1936 428}%
\special{pa 1940 432}%
\special{pa 1946 434}%
\special{pa 1950 434}%
\special{pa 1956 434}%
\special{pa 1960 434}%
\special{pa 1966 430}%
\special{pa 1970 428}%
\special{pa 1976 424}%
\special{pa 1980 420}%
\special{pa 1986 414}%
\special{pa 1990 410}%
\special{pa 1996 404}%
\special{pa 2000 400}%
\special{pa 2006 396}%
\special{pa 2010 392}%
\special{pa 2016 390}%
\special{pa 2020 388}%
\special{pa 2026 386}%
\special{pa 2030 386}%
\special{pa 2036 388}%
\special{pa 2040 390}%
\special{pa 2046 394}%
\special{pa 2050 398}%
\special{pa 2056 402}%
\special{pa 2060 408}%
\special{pa 2066 412}%
\special{pa 2070 418}%
\special{pa 2076 422}%
\special{pa 2080 426}%
\special{pa 2086 430}%
\special{pa 2090 432}%
\special{pa 2096 434}%
\special{pa 2100 434}%
\special{pa 2106 434}%
\special{pa 2110 432}%
\special{pa 2116 430}%
\special{pa 2120 426}%
\special{pa 2126 422}%
\special{pa 2130 418}%
\special{pa 2136 412}%
\special{pa 2140 408}%
\special{pa 2146 402}%
\special{pa 2150 398}%
\special{pa 2156 394}%
\special{pa 2160 390}%
\special{pa 2166 388}%
\special{pa 2170 386}%
\special{pa 2176 386}%
\special{pa 2180 388}%
\special{pa 2186 390}%
\special{pa 2190 392}%
\special{pa 2196 396}%
\special{pa 2200 400}%
\special{pa 2206 404}%
\special{pa 2210 410}%
\special{pa 2216 416}%
\special{pa 2220 420}%
\special{pa 2226 424}%
\special{pa 2230 428}%
\special{pa 2236 432}%
\special{pa 2240 434}%
\special{pa 2246 434}%
\special{pa 2250 434}%
\special{pa 2256 434}%
\special{pa 2260 432}%
\special{pa 2266 428}%
\special{pa 2270 424}%
\special{pa 2276 420}%
\special{pa 2280 414}%
\special{pa 2286 410}%
\special{pa 2290 404}%
\special{pa 2296 400}%
\special{pa 2300 396}%
\special{pa 2306 392}%
\special{pa 2310 390}%
\special{pa 2316 388}%
\special{pa 2320 386}%
\special{pa 2326 386}%
\special{pa 2330 388}%
\special{pa 2336 390}%
\special{pa 2340 394}%
\special{pa 2346 398}%
\special{pa 2350 402}%
\special{pa 2356 408}%
\special{pa 2360 412}%
\special{pa 2366 418}%
\special{pa 2370 422}%
\special{pa 2376 426}%
\special{pa 2380 430}%
\special{pa 2386 432}%
\special{pa 2390 434}%
\special{pa 2396 434}%
\special{pa 2400 434}%
\special{pa 2406 432}%
\special{pa 2410 430}%
\special{pa 2416 426}%
\special{pa 2420 422}%
\special{pa 2426 418}%
\special{pa 2430 412}%
\special{pa 2436 408}%
\special{pa 2440 402}%
\special{pa 2446 398}%
\special{pa 2450 394}%
\special{pa 2456 390}%
\special{pa 2460 388}%
\special{pa 2466 386}%
\special{pa 2470 386}%
\special{pa 2476 388}%
\special{pa 2480 390}%
\special{pa 2486 392}%
\special{pa 2490 396}%
\special{sp}%
\special{pn 13}%
\special{pa 1546 412}%
\special{pa 1550 406}%
\special{pa 1556 402}%
\special{pa 1560 398}%
\special{pa 1566 394}%
\special{pa 1570 390}%
\special{pa 1576 388}%
\special{pa 1580 386}%
\special{pa 1586 386}%
\special{pa 1590 388}%
\special{pa 1596 390}%
\special{pa 1600 392}%
\special{pa 1606 396}%
\special{pa 1610 400}%
\special{pa 1616 406}%
\special{pa 1620 410}%
\special{pa 1626 416}%
\special{pa 1630 420}%
\special{pa 1636 424}%
\special{pa 1640 428}%
\special{pa 1646 432}%
\special{pa 1650 434}%
\special{pa 1656 434}%
\special{pa 1660 434}%
\special{pa 1666 434}%
\special{pa 1670 430}%
\special{pa 1676 428}%
\special{pa 1680 424}%
\special{pa 1686 420}%
\special{pa 1690 414}%
\special{pa 1696 410}%
\special{pa 1700 404}%
\special{pa 1706 400}%
\special{pa 1710 396}%
\special{pa 1716 392}%
\special{pa 1720 388}%
\special{pa 1726 388}%
\special{pa 1730 386}%
\special{pa 1736 386}%
\special{pa 1740 388}%
\special{pa 1746 390}%
\special{pa 1750 394}%
\special{pa 1756 398}%
\special{pa 1760 402}%
\special{pa 1766 408}%
\special{pa 1770 414}%
\special{pa 1776 418}%
\special{pa 1780 422}%
\special{pa 1786 426}%
\special{pa 1790 430}%
\special{pa 1796 432}%
\special{pa 1800 434}%
\special{pa 1806 434}%
\special{pa 1810 434}%
\special{pa 1816 432}%
\special{pa 1820 430}%
\special{pa 1826 426}%
\special{pa 1830 422}%
\special{pa 1836 416}%
\special{pa 1840 412}%
\special{pa 1846 406}%
\special{pa 1850 402}%
\special{pa 1856 398}%
\special{pa 1860 394}%
\special{pa 1866 390}%
\special{pa 1870 388}%
\special{pa 1876 386}%
\special{pa 1880 386}%
\special{pa 1886 388}%
\special{pa 1890 390}%
\special{pa 1896 392}%
\special{pa 1900 396}%
\special{pa 1906 400}%
\special{pa 1910 406}%
\special{pa 1916 410}%
\special{pa 1920 416}%
\special{pa 1926 420}%
\special{pa 1930 424}%
\special{pa 1936 428}%
\special{pa 1940 432}%
\special{pa 1946 434}%
\special{pa 1950 434}%
\special{pa 1956 434}%
\special{pa 1960 434}%
\special{pa 1966 430}%
\special{pa 1970 428}%
\special{pa 1976 424}%
\special{pa 1980 420}%
\special{pa 1986 414}%
\special{pa 1990 410}%
\special{pa 1996 404}%
\special{pa 2000 400}%
\special{pa 2006 396}%
\special{pa 2010 392}%
\special{pa 2016 390}%
\special{pa 2020 388}%
\special{pa 2026 386}%
\special{pa 2030 386}%
\special{pa 2036 388}%
\special{pa 2040 390}%
\special{pa 2046 394}%
\special{pa 2050 398}%
\special{pa 2056 402}%
\special{pa 2060 408}%
\special{pa 2066 412}%
\special{pa 2070 418}%
\special{pa 2076 422}%
\special{pa 2080 426}%
\special{pa 2086 430}%
\special{pa 2090 432}%
\special{pa 2096 434}%
\special{pa 2100 434}%
\special{pa 2106 434}%
\special{pa 2110 432}%
\special{pa 2116 430}%
\special{pa 2120 426}%
\special{pa 2126 422}%
\special{pa 2130 418}%
\special{pa 2136 412}%
\special{pa 2140 408}%
\special{pa 2146 402}%
\special{pa 2150 398}%
\special{pa 2156 394}%
\special{pa 2160 390}%
\special{pa 2166 388}%
\special{pa 2170 386}%
\special{pa 2176 386}%
\special{pa 2180 388}%
\special{pa 2186 390}%
\special{pa 2190 392}%
\special{pa 2196 396}%
\special{pa 2200 400}%
\special{pa 2206 404}%
\special{pa 2210 410}%
\special{pa 2216 416}%
\special{pa 2220 420}%
\special{pa 2226 424}%
\special{pa 2230 428}%
\special{pa 2236 432}%
\special{pa 2240 434}%
\special{pa 2246 434}%
\special{pa 2250 434}%
\special{pa 2256 434}%
\special{pa 2260 432}%
\special{pa 2266 428}%
\special{pa 2270 424}%
\special{pa 2276 420}%
\special{pa 2280 414}%
\special{pa 2286 410}%
\special{pa 2290 404}%
\special{pa 2296 400}%
\special{pa 2300 396}%
\special{pa 2306 392}%
\special{pa 2310 390}%
\special{pa 2316 388}%
\special{pa 2320 386}%
\special{pa 2326 386}%
\special{pa 2330 388}%
\special{pa 2336 390}%
\special{pa 2340 394}%
\special{pa 2346 398}%
\special{pa 2350 402}%
\special{pa 2356 408}%
\special{pa 2360 412}%
\special{pa 2366 418}%
\special{pa 2370 422}%
\special{pa 2376 426}%
\special{pa 2380 430}%
\special{pa 2386 432}%
\special{pa 2390 434}%
\special{pa 2396 434}%
\special{pa 2400 434}%
\special{pa 2406 432}%
\special{pa 2410 430}%
\special{pa 2416 426}%
\special{pa 2420 422}%
\special{pa 2426 418}%
\special{pa 2430 412}%
\special{pa 2436 408}%
\special{pa 2440 402}%
\special{pa 2446 398}%
\special{pa 2450 394}%
\special{pa 2456 390}%
\special{pa 2460 388}%
\special{pa 2466 386}%
\special{pa 2470 386}%
\special{pa 2476 388}%
\special{pa 2480 390}%
\special{pa 2486 392}%
\special{pa 2490 396}%
\special{sp}%
\special{pn 13}%
\special{pa 1546 1012}%
\special{pa 1550 1006}%
\special{pa 1556 1002}%
\special{pa 1560 998}%
\special{pa 1566 994}%
\special{pa 1570 990}%
\special{pa 1576 988}%
\special{pa 1580 986}%
\special{pa 1586 986}%
\special{pa 1590 988}%
\special{pa 1596 990}%
\special{pa 1600 992}%
\special{pa 1606 996}%
\special{pa 1610 1000}%
\special{pa 1616 1006}%
\special{pa 1620 1010}%
\special{pa 1626 1016}%
\special{pa 1630 1020}%
\special{pa 1636 1024}%
\special{pa 1640 1028}%
\special{pa 1646 1032}%
\special{pa 1650 1034}%
\special{pa 1656 1034}%
\special{pa 1660 1034}%
\special{pa 1666 1034}%
\special{pa 1670 1030}%
\special{pa 1676 1028}%
\special{pa 1680 1024}%
\special{pa 1686 1020}%
\special{pa 1690 1014}%
\special{pa 1696 1010}%
\special{pa 1700 1004}%
\special{pa 1706 1000}%
\special{pa 1710 996}%
\special{pa 1716 992}%
\special{pa 1720 988}%
\special{pa 1726 988}%
\special{pa 1730 986}%
\special{pa 1736 986}%
\special{pa 1740 988}%
\special{pa 1746 990}%
\special{pa 1750 994}%
\special{pa 1756 998}%
\special{pa 1760 1002}%
\special{pa 1766 1008}%
\special{pa 1770 1014}%
\special{pa 1776 1018}%
\special{pa 1780 1022}%
\special{pa 1786 1026}%
\special{pa 1790 1030}%
\special{pa 1796 1032}%
\special{pa 1800 1034}%
\special{pa 1806 1034}%
\special{pa 1810 1034}%
\special{pa 1816 1032}%
\special{pa 1820 1030}%
\special{pa 1826 1026}%
\special{pa 1830 1022}%
\special{pa 1836 1016}%
\special{pa 1840 1012}%
\special{pa 1846 1006}%
\special{pa 1850 1002}%
\special{pa 1856 998}%
\special{pa 1860 994}%
\special{pa 1866 990}%
\special{pa 1870 988}%
\special{pa 1876 986}%
\special{pa 1880 986}%
\special{pa 1886 988}%
\special{pa 1890 990}%
\special{pa 1896 992}%
\special{pa 1900 996}%
\special{pa 1906 1000}%
\special{pa 1910 1006}%
\special{pa 1916 1010}%
\special{pa 1920 1016}%
\special{pa 1926 1020}%
\special{pa 1930 1024}%
\special{pa 1936 1028}%
\special{pa 1940 1032}%
\special{pa 1946 1034}%
\special{pa 1950 1034}%
\special{pa 1956 1034}%
\special{pa 1960 1034}%
\special{pa 1966 1030}%
\special{pa 1970 1028}%
\special{pa 1976 1024}%
\special{pa 1980 1020}%
\special{pa 1986 1014}%
\special{pa 1990 1010}%
\special{pa 1996 1004}%
\special{pa 2000 1000}%
\special{pa 2006 996}%
\special{pa 2010 992}%
\special{pa 2016 990}%
\special{pa 2020 988}%
\special{pa 2026 986}%
\special{pa 2030 986}%
\special{pa 2036 988}%
\special{pa 2040 990}%
\special{pa 2046 994}%
\special{pa 2050 998}%
\special{pa 2056 1002}%
\special{pa 2060 1008}%
\special{pa 2066 1012}%
\special{pa 2070 1018}%
\special{pa 2076 1022}%
\special{pa 2080 1026}%
\special{pa 2086 1030}%
\special{pa 2090 1032}%
\special{pa 2096 1034}%
\special{pa 2100 1034}%
\special{pa 2106 1034}%
\special{pa 2110 1032}%
\special{pa 2116 1030}%
\special{pa 2120 1026}%
\special{pa 2126 1022}%
\special{pa 2130 1018}%
\special{pa 2136 1012}%
\special{pa 2140 1008}%
\special{pa 2146 1002}%
\special{pa 2150 998}%
\special{pa 2156 994}%
\special{pa 2160 990}%
\special{pa 2166 988}%
\special{pa 2170 986}%
\special{pa 2176 986}%
\special{pa 2180 988}%
\special{pa 2186 990}%
\special{pa 2190 992}%
\special{pa 2196 996}%
\special{pa 2200 1000}%
\special{pa 2206 1004}%
\special{pa 2210 1010}%
\special{pa 2216 1016}%
\special{pa 2220 1020}%
\special{pa 2226 1024}%
\special{pa 2230 1028}%
\special{pa 2236 1032}%
\special{pa 2240 1034}%
\special{pa 2246 1034}%
\special{pa 2250 1034}%
\special{pa 2256 1034}%
\special{pa 2260 1032}%
\special{pa 2266 1028}%
\special{pa 2270 1024}%
\special{pa 2276 1020}%
\special{pa 2280 1014}%
\special{pa 2286 1010}%
\special{pa 2290 1004}%
\special{pa 2296 1000}%
\special{pa 2300 996}%
\special{pa 2306 992}%
\special{pa 2310 990}%
\special{pa 2316 988}%
\special{pa 2320 986}%
\special{pa 2326 986}%
\special{pa 2330 988}%
\special{pa 2336 990}%
\special{pa 2340 994}%
\special{pa 2346 998}%
\special{pa 2350 1002}%
\special{pa 2356 1008}%
\special{pa 2360 1012}%
\special{pa 2366 1018}%
\special{pa 2370 1022}%
\special{pa 2376 1026}%
\special{pa 2380 1030}%
\special{pa 2386 1032}%
\special{pa 2390 1034}%
\special{pa 2396 1034}%
\special{pa 2400 1034}%
\special{pa 2406 1032}%
\special{pa 2410 1030}%
\special{pa 2416 1026}%
\special{pa 2420 1022}%
\special{pa 2426 1018}%
\special{pa 2430 1012}%
\special{pa 2436 1008}%
\special{pa 2440 1002}%
\special{pa 2446 998}%
\special{pa 2450 994}%
\special{pa 2456 990}%
\special{pa 2460 988}%
\special{pa 2466 986}%
\special{pa 2470 986}%
\special{pa 2476 988}%
\special{pa 2480 990}%
\special{pa 2486 992}%
\special{pa 2490 996}%
\special{sp}%
%
\special{pn 8}%
\special{pa 1310 410}%
\special{pa 1310 1010}%
\special{fp}%
\special{pa 1782 1010}%
\special{pa 1782 410}%
\special{fp}%
%
\special{pn 8}%
\special{pa 1782 410}%
\special{pa 1782 1010}%
\special{fp}%
%
\special{pn 8}%
\special{pa 1782 410}%
\special{pa 2490 410}%
\special{pa 2490 1010}%
\special{pa 1782 1010}%
\special{pa 1782 410}%
\special{ip}%
\put(26.0000,-7.5000){\makebox(0,0)[lb]{$gg$}}%
\special{pn 8}%
\special{pa 4286 412}%
\special{pa 4290 406}%
\special{pa 4296 402}%
\special{pa 4300 398}%
\special{pa 4306 394}%
\special{pa 4310 390}%
\special{pa 4316 388}%
\special{pa 4320 386}%
\special{pa 4326 386}%
\special{pa 4330 388}%
\special{pa 4336 390}%
\special{pa 4340 392}%
\special{pa 4346 396}%
\special{pa 4350 400}%
\special{pa 4356 406}%
\special{pa 4360 410}%
\special{pa 4366 416}%
\special{pa 4370 420}%
\special{pa 4376 424}%
\special{pa 4380 428}%
\special{pa 4386 432}%
\special{pa 4390 434}%
\special{pa 4396 434}%
\special{pa 4400 434}%
\special{pa 4406 434}%
\special{pa 4410 430}%
\special{pa 4416 428}%
\special{pa 4420 424}%
\special{pa 4426 420}%
\special{pa 4430 414}%
\special{pa 4436 410}%
\special{pa 4440 404}%
\special{pa 4446 400}%
\special{pa 4450 396}%
\special{pa 4456 392}%
\special{pa 4460 388}%
\special{pa 4466 388}%
\special{pa 4470 386}%
\special{pa 4476 386}%
\special{pa 4480 388}%
\special{pa 4486 390}%
\special{pa 4490 394}%
\special{pa 4496 398}%
\special{pa 4500 402}%
\special{pa 4506 408}%
\special{pa 4510 414}%
\special{pa 4516 418}%
\special{pa 4520 422}%
\special{pa 4526 426}%
\special{pa 4530 430}%
\special{pa 4536 432}%
\special{pa 4540 434}%
\special{pa 4546 434}%
\special{pa 4550 434}%
\special{pa 4556 432}%
\special{pa 4560 430}%
\special{pa 4566 426}%
\special{pa 4570 422}%
\special{pa 4576 416}%
\special{pa 4580 412}%
\special{pa 4586 406}%
\special{pa 4590 402}%
\special{pa 4596 398}%
\special{pa 4600 394}%
\special{pa 4606 390}%
\special{pa 4610 388}%
\special{pa 4616 386}%
\special{pa 4620 386}%
\special{pa 4626 388}%
\special{pa 4630 390}%
\special{pa 4636 392}%
\special{pa 4640 396}%
\special{pa 4646 400}%
\special{pa 4650 406}%
\special{pa 4656 410}%
\special{pa 4660 416}%
\special{pa 4666 420}%
\special{pa 4670 424}%
\special{pa 4676 428}%
\special{pa 4680 432}%
\special{pa 4686 434}%
\special{pa 4690 434}%
\special{pa 4696 434}%
\special{pa 4700 434}%
\special{pa 4706 430}%
\special{pa 4710 428}%
\special{pa 4716 424}%
\special{pa 4720 420}%
\special{pa 4726 414}%
\special{pa 4730 410}%
\special{pa 4736 404}%
\special{pa 4740 400}%
\special{pa 4746 396}%
\special{pa 4750 392}%
\special{pa 4756 390}%
\special{pa 4760 388}%
\special{pa 4766 386}%
\special{pa 4770 386}%
\special{pa 4776 388}%
\special{pa 4780 390}%
\special{pa 4786 394}%
\special{pa 4790 398}%
\special{pa 4796 402}%
\special{pa 4800 408}%
\special{pa 4806 412}%
\special{pa 4810 418}%
\special{pa 4816 422}%
\special{pa 4820 426}%
\special{pa 4826 430}%
\special{pa 4830 432}%
\special{pa 4836 434}%
\special{pa 4840 434}%
\special{pa 4846 434}%
\special{pa 4850 432}%
\special{pa 4856 430}%
\special{pa 4860 426}%
\special{pa 4866 422}%
\special{pa 4870 418}%
\special{pa 4876 412}%
\special{pa 4880 408}%
\special{pa 4886 402}%
\special{pa 4890 398}%
\special{pa 4896 394}%
\special{pa 4900 390}%
\special{pa 4906 388}%
\special{pa 4910 386}%
\special{pa 4916 386}%
\special{pa 4920 388}%
\special{pa 4926 390}%
\special{pa 4930 392}%
\special{pa 4936 396}%
\special{pa 4940 400}%
\special{pa 4946 404}%
\special{pa 4950 410}%
\special{pa 4956 416}%
\special{pa 4960 420}%
\special{pa 4966 424}%
\special{pa 4970 428}%
\special{pa 4976 432}%
\special{pa 4980 434}%
\special{pa 4986 434}%
\special{pa 4990 434}%
\special{pa 4996 434}%
\special{pa 5000 432}%
\special{pa 5006 428}%
\special{pa 5010 424}%
\special{pa 5016 420}%
\special{pa 5020 414}%
\special{pa 5026 410}%
\special{pa 5030 404}%
\special{pa 5036 400}%
\special{pa 5040 396}%
\special{pa 5046 392}%
\special{pa 5050 390}%
\special{pa 5056 388}%
\special{pa 5060 386}%
\special{pa 5066 386}%
\special{pa 5070 388}%
\special{pa 5076 390}%
\special{pa 5080 394}%
\special{pa 5086 398}%
\special{pa 5090 402}%
\special{pa 5096 408}%
\special{pa 5100 412}%
\special{pa 5106 418}%
\special{pa 5110 422}%
\special{pa 5116 426}%
\special{pa 5120 430}%
\special{pa 5126 432}%
\special{pa 5130 434}%
\special{pa 5136 434}%
\special{pa 5140 434}%
\special{pa 5146 432}%
\special{pa 5150 430}%
\special{pa 5156 426}%
\special{pa 5160 422}%
\special{pa 5166 418}%
\special{pa 5170 412}%
\special{pa 5176 408}%
\special{pa 5180 402}%
\special{pa 5186 398}%
\special{pa 5190 394}%
\special{pa 5196 390}%
\special{pa 5200 388}%
\special{pa 5206 386}%
\special{pa 5210 386}%
\special{pa 5216 388}%
\special{pa 5220 390}%
\special{pa 5226 392}%
\special{pa 5230 396}%
\special{sp}%
\special{pn 13}%
\special{pa 4286 412}%
\special{pa 4290 406}%
\special{pa 4296 402}%
\special{pa 4300 398}%
\special{pa 4306 394}%
\special{pa 4310 390}%
\special{pa 4316 388}%
\special{pa 4320 386}%
\special{pa 4326 386}%
\special{pa 4330 388}%
\special{pa 4336 390}%
\special{pa 4340 392}%
\special{pa 4346 396}%
\special{pa 4350 400}%
\special{pa 4356 406}%
\special{pa 4360 410}%
\special{pa 4366 416}%
\special{pa 4370 420}%
\special{pa 4376 424}%
\special{pa 4380 428}%
\special{pa 4386 432}%
\special{pa 4390 434}%
\special{pa 4396 434}%
\special{pa 4400 434}%
\special{pa 4406 434}%
\special{pa 4410 430}%
\special{pa 4416 428}%
\special{pa 4420 424}%
\special{pa 4426 420}%
\special{pa 4430 414}%
\special{pa 4436 410}%
\special{pa 4440 404}%
\special{pa 4446 400}%
\special{pa 4450 396}%
\special{pa 4456 392}%
\special{pa 4460 388}%
\special{pa 4466 388}%
\special{pa 4470 386}%
\special{pa 4476 386}%
\special{pa 4480 388}%
\special{pa 4486 390}%
\special{pa 4490 394}%
\special{pa 4496 398}%
\special{pa 4500 402}%
\special{pa 4506 408}%
\special{pa 4510 414}%
\special{pa 4516 418}%
\special{pa 4520 422}%
\special{pa 4526 426}%
\special{pa 4530 430}%
\special{pa 4536 432}%
\special{pa 4540 434}%
\special{pa 4546 434}%
\special{pa 4550 434}%
\special{pa 4556 432}%
\special{pa 4560 430}%
\special{pa 4566 426}%
\special{pa 4570 422}%
\special{pa 4576 416}%
\special{pa 4580 412}%
\special{pa 4586 406}%
\special{pa 4590 402}%
\special{pa 4596 398}%
\special{pa 4600 394}%
\special{pa 4606 390}%
\special{pa 4610 388}%
\special{pa 4616 386}%
\special{pa 4620 386}%
\special{pa 4626 388}%
\special{pa 4630 390}%
\special{pa 4636 392}%
\special{pa 4640 396}%
\special{pa 4646 400}%
\special{pa 4650 406}%
\special{pa 4656 410}%
\special{pa 4660 416}%
\special{pa 4666 420}%
\special{pa 4670 424}%
\special{pa 4676 428}%
\special{pa 4680 432}%
\special{pa 4686 434}%
\special{pa 4690 434}%
\special{pa 4696 434}%
\special{pa 4700 434}%
\special{pa 4706 430}%
\special{pa 4710 428}%
\special{pa 4716 424}%
\special{pa 4720 420}%
\special{pa 4726 414}%
\special{pa 4730 410}%
\special{pa 4736 404}%
\special{pa 4740 400}%
\special{pa 4746 396}%
\special{pa 4750 392}%
\special{pa 4756 390}%
\special{pa 4760 388}%
\special{pa 4766 386}%
\special{pa 4770 386}%
\special{pa 4776 388}%
\special{pa 4780 390}%
\special{pa 4786 394}%
\special{pa 4790 398}%
\special{pa 4796 402}%
\special{pa 4800 408}%
\special{pa 4806 412}%
\special{pa 4810 418}%
\special{pa 4816 422}%
\special{pa 4820 426}%
\special{pa 4826 430}%
\special{pa 4830 432}%
\special{pa 4836 434}%
\special{pa 4840 434}%
\special{pa 4846 434}%
\special{pa 4850 432}%
\special{pa 4856 430}%
\special{pa 4860 426}%
\special{pa 4866 422}%
\special{pa 4870 418}%
\special{pa 4876 412}%
\special{pa 4880 408}%
\special{pa 4886 402}%
\special{pa 4890 398}%
\special{pa 4896 394}%
\special{pa 4900 390}%
\special{pa 4906 388}%
\special{pa 4910 386}%
\special{pa 4916 386}%
\special{pa 4920 388}%
\special{pa 4926 390}%
\special{pa 4930 392}%
\special{pa 4936 396}%
\special{pa 4940 400}%
\special{pa 4946 404}%
\special{pa 4950 410}%
\special{pa 4956 416}%
\special{pa 4960 420}%
\special{pa 4966 424}%
\special{pa 4970 428}%
\special{pa 4976 432}%
\special{pa 4980 434}%
\special{pa 4986 434}%
\special{pa 4990 434}%
\special{pa 4996 434}%
\special{pa 5000 432}%
\special{pa 5006 428}%
\special{pa 5010 424}%
\special{pa 5016 420}%
\special{pa 5020 414}%
\special{pa 5026 410}%
\special{pa 5030 404}%
\special{pa 5036 400}%
\special{pa 5040 396}%
\special{pa 5046 392}%
\special{pa 5050 390}%
\special{pa 5056 388}%
\special{pa 5060 386}%
\special{pa 5066 386}%
\special{pa 5070 388}%
\special{pa 5076 390}%
\special{pa 5080 394}%
\special{pa 5086 398}%
\special{pa 5090 402}%
\special{pa 5096 408}%
\special{pa 5100 412}%
\special{pa 5106 418}%
\special{pa 5110 422}%
\special{pa 5116 426}%
\special{pa 5120 430}%
\special{pa 5126 432}%
\special{pa 5130 434}%
\special{pa 5136 434}%
\special{pa 5140 434}%
\special{pa 5146 432}%
\special{pa 5150 430}%
\special{pa 5156 426}%
\special{pa 5160 422}%
\special{pa 5166 418}%
\special{pa 5170 412}%
\special{pa 5176 408}%
\special{pa 5180 402}%
\special{pa 5186 398}%
\special{pa 5190 394}%
\special{pa 5196 390}%
\special{pa 5200 388}%
\special{pa 5206 386}%
\special{pa 5210 386}%
\special{pa 5216 388}%
\special{pa 5220 390}%
\special{pa 5226 392}%
\special{pa 5230 396}%
\special{sp}%
\special{pn 13}%
\special{pa 4286 1012}%
\special{pa 4290 1006}%
\special{pa 4296 1002}%
\special{pa 4300 998}%
\special{pa 4306 994}%
\special{pa 4310 990}%
\special{pa 4316 988}%
\special{pa 4320 986}%
\special{pa 4326 986}%
\special{pa 4330 988}%
\special{pa 4336 990}%
\special{pa 4340 992}%
\special{pa 4346 996}%
\special{pa 4350 1000}%
\special{pa 4356 1006}%
\special{pa 4360 1010}%
\special{pa 4366 1016}%
\special{pa 4370 1020}%
\special{pa 4376 1024}%
\special{pa 4380 1028}%
\special{pa 4386 1032}%
\special{pa 4390 1034}%
\special{pa 4396 1034}%
\special{pa 4400 1034}%
\special{pa 4406 1034}%
\special{pa 4410 1030}%
\special{pa 4416 1028}%
\special{pa 4420 1024}%
\special{pa 4426 1020}%
\special{pa 4430 1014}%
\special{pa 4436 1010}%
\special{pa 4440 1004}%
\special{pa 4446 1000}%
\special{pa 4450 996}%
\special{pa 4456 992}%
\special{pa 4460 988}%
\special{pa 4466 988}%
\special{pa 4470 986}%
\special{pa 4476 986}%
\special{pa 4480 988}%
\special{pa 4486 990}%
\special{pa 4490 994}%
\special{pa 4496 998}%
\special{pa 4500 1002}%
\special{pa 4506 1008}%
\special{pa 4510 1014}%
\special{pa 4516 1018}%
\special{pa 4520 1022}%
\special{pa 4526 1026}%
\special{pa 4530 1030}%
\special{pa 4536 1032}%
\special{pa 4540 1034}%
\special{pa 4546 1034}%
\special{pa 4550 1034}%
\special{pa 4556 1032}%
\special{pa 4560 1030}%
\special{pa 4566 1026}%
\special{pa 4570 1022}%
\special{pa 4576 1016}%
\special{pa 4580 1012}%
\special{pa 4586 1006}%
\special{pa 4590 1002}%
\special{pa 4596 998}%
\special{pa 4600 994}%
\special{pa 4606 990}%
\special{pa 4610 988}%
\special{pa 4616 986}%
\special{pa 4620 986}%
\special{pa 4626 988}%
\special{pa 4630 990}%
\special{pa 4636 992}%
\special{pa 4640 996}%
\special{pa 4646 1000}%
\special{pa 4650 1006}%
\special{pa 4656 1010}%
\special{pa 4660 1016}%
\special{pa 4666 1020}%
\special{pa 4670 1024}%
\special{pa 4676 1028}%
\special{pa 4680 1032}%
\special{pa 4686 1034}%
\special{pa 4690 1034}%
\special{pa 4696 1034}%
\special{pa 4700 1034}%
\special{pa 4706 1030}%
\special{pa 4710 1028}%
\special{pa 4716 1024}%
\special{pa 4720 1020}%
\special{pa 4726 1014}%
\special{pa 4730 1010}%
\special{pa 4736 1004}%
\special{pa 4740 1000}%
\special{pa 4746 996}%
\special{pa 4750 992}%
\special{pa 4756 990}%
\special{pa 4760 988}%
\special{pa 4766 986}%
\special{pa 4770 986}%
\special{pa 4776 988}%
\special{pa 4780 990}%
\special{pa 4786 994}%
\special{pa 4790 998}%
\special{pa 4796 1002}%
\special{pa 4800 1008}%
\special{pa 4806 1012}%
\special{pa 4810 1018}%
\special{pa 4816 1022}%
\special{pa 4820 1026}%
\special{pa 4826 1030}%
\special{pa 4830 1032}%
\special{pa 4836 1034}%
\special{pa 4840 1034}%
\special{pa 4846 1034}%
\special{pa 4850 1032}%
\special{pa 4856 1030}%
\special{pa 4860 1026}%
\special{pa 4866 1022}%
\special{pa 4870 1018}%
\special{pa 4876 1012}%
\special{pa 4880 1008}%
\special{pa 4886 1002}%
\special{pa 4890 998}%
\special{pa 4896 994}%
\special{pa 4900 990}%
\special{pa 4906 988}%
\special{pa 4910 986}%
\special{pa 4916 986}%
\special{pa 4920 988}%
\special{pa 4926 990}%
\special{pa 4930 992}%
\special{pa 4936 996}%
\special{pa 4940 1000}%
\special{pa 4946 1004}%
\special{pa 4950 1010}%
\special{pa 4956 1016}%
\special{pa 4960 1020}%
\special{pa 4966 1024}%
\special{pa 4970 1028}%
\special{pa 4976 1032}%
\special{pa 4980 1034}%
\special{pa 4986 1034}%
\special{pa 4990 1034}%
\special{pa 4996 1034}%
\special{pa 5000 1032}%
\special{pa 5006 1028}%
\special{pa 5010 1024}%
\special{pa 5016 1020}%
\special{pa 5020 1014}%
\special{pa 5026 1010}%
\special{pa 5030 1004}%
\special{pa 5036 1000}%
\special{pa 5040 996}%
\special{pa 5046 992}%
\special{pa 5050 990}%
\special{pa 5056 988}%
\special{pa 5060 986}%
\special{pa 5066 986}%
\special{pa 5070 988}%
\special{pa 5076 990}%
\special{pa 5080 994}%
\special{pa 5086 998}%
\special{pa 5090 1002}%
\special{pa 5096 1008}%
\special{pa 5100 1012}%
\special{pa 5106 1018}%
\special{pa 5110 1022}%
\special{pa 5116 1026}%
\special{pa 5120 1030}%
\special{pa 5126 1032}%
\special{pa 5130 1034}%
\special{pa 5136 1034}%
\special{pa 5140 1034}%
\special{pa 5146 1032}%
\special{pa 5150 1030}%
\special{pa 5156 1026}%
\special{pa 5160 1022}%
\special{pa 5166 1018}%
\special{pa 5170 1012}%
\special{pa 5176 1008}%
\special{pa 5180 1002}%
\special{pa 5186 998}%
\special{pa 5190 994}%
\special{pa 5196 990}%
\special{pa 5200 988}%
\special{pa 5206 986}%
\special{pa 5210 986}%
\special{pa 5216 988}%
\special{pa 5220 990}%
\special{pa 5226 992}%
\special{pa 5230 996}%
\special{sp}%
%
\special{pn 8}%
\special{pa 4522 410}%
\special{pa 4522 1010}%
\special{fp}%
%
\special{pn 8}%
\special{pa 4522 410}%
\special{pa 5230 410}%
\special{pa 5230 1010}%
\special{pa 4522 1010}%
\special{pa 4522 410}%
\special{ip}%
\put(53.4000,-7.5000){\makebox(0,0)[lb]{$gg$}}%
\special{pn 8}%
\special{pa 3390 410}%
\special{pa 3396 406}%
\special{pa 3400 400}%
\special{pa 3406 396}%
\special{pa 3410 392}%
\special{pa 3416 390}%
\special{pa 3420 388}%
\special{pa 3426 386}%
\special{pa 3430 386}%
\special{pa 3436 388}%
\special{pa 3440 390}%
\special{pa 3446 394}%
\special{pa 3450 398}%
\special{pa 3456 402}%
\special{pa 3460 406}%
\special{pa 3466 412}%
\special{pa 3470 416}%
\special{pa 3476 422}%
\special{pa 3480 426}%
\special{pa 3486 430}%
\special{pa 3490 432}%
\special{pa 3496 434}%
\special{pa 3500 434}%
\special{pa 3506 434}%
\special{pa 3510 432}%
\special{pa 3516 430}%
\special{pa 3520 426}%
\special{pa 3526 422}%
\special{pa 3530 418}%
\special{pa 3536 414}%
\special{pa 3540 408}%
\special{pa 3546 404}%
\special{pa 3550 398}%
\special{pa 3556 394}%
\special{pa 3560 390}%
\special{pa 3566 388}%
\special{pa 3570 386}%
\special{pa 3576 386}%
\special{pa 3580 388}%
\special{pa 3586 388}%
\special{pa 3590 392}%
\special{pa 3596 396}%
\special{pa 3600 400}%
\special{pa 3606 404}%
\special{pa 3610 408}%
\special{pa 3616 414}%
\special{pa 3620 420}%
\special{pa 3626 424}%
\special{pa 3630 428}%
\special{pa 3636 430}%
\special{pa 3640 432}%
\special{pa 3646 434}%
\special{pa 3650 434}%
\special{pa 3656 434}%
\special{pa 3660 432}%
\special{pa 3666 428}%
\special{pa 3670 426}%
\special{pa 3676 420}%
\special{pa 3680 416}%
\special{pa 3686 410}%
\special{pa 3690 406}%
\special{pa 3696 400}%
\special{pa 3700 396}%
\special{pa 3706 392}%
\special{pa 3710 390}%
\special{pa 3716 388}%
\special{pa 3720 386}%
\special{pa 3726 386}%
\special{pa 3730 388}%
\special{pa 3736 390}%
\special{pa 3740 394}%
\special{pa 3746 396}%
\special{pa 3750 402}%
\special{pa 3756 406}%
\special{pa 3760 412}%
\special{pa 3766 416}%
\special{pa 3770 422}%
\special{pa 3776 426}%
\special{pa 3780 430}%
\special{pa 3786 432}%
\special{pa 3790 434}%
\special{pa 3796 434}%
\special{pa 3800 434}%
\special{pa 3806 432}%
\special{pa 3810 430}%
\special{pa 3816 428}%
\special{pa 3820 424}%
\special{pa 3826 418}%
\special{pa 3830 414}%
\special{pa 3836 408}%
\special{pa 3840 404}%
\special{pa 3846 398}%
\special{pa 3850 394}%
\special{pa 3856 392}%
\special{pa 3860 388}%
\special{pa 3866 388}%
\special{pa 3870 386}%
\special{pa 3876 388}%
\special{pa 3880 388}%
\special{pa 3886 392}%
\special{pa 3890 394}%
\special{pa 3896 398}%
\special{pa 3900 404}%
\special{pa 3906 408}%
\special{pa 3910 414}%
\special{pa 3916 418}%
\special{pa 3920 424}%
\special{pa 3926 428}%
\special{pa 3930 430}%
\special{pa 3936 432}%
\special{pa 3940 434}%
\special{pa 3946 434}%
\special{pa 3950 434}%
\special{pa 3956 432}%
\special{pa 3960 428}%
\special{pa 3966 426}%
\special{pa 3970 420}%
\special{pa 3976 416}%
\special{pa 3980 412}%
\special{pa 3986 406}%
\special{pa 3990 402}%
\special{pa 3996 396}%
\special{pa 4000 392}%
\special{pa 4006 390}%
\special{pa 4010 388}%
\special{pa 4016 386}%
\special{pa 4020 386}%
\special{pa 4026 388}%
\special{pa 4030 390}%
\special{pa 4036 392}%
\special{pa 4040 396}%
\special{pa 4046 402}%
\special{pa 4050 406}%
\special{pa 4056 412}%
\special{pa 4060 416}%
\special{pa 4066 420}%
\special{pa 4070 426}%
\special{pa 4076 428}%
\special{pa 4080 432}%
\special{pa 4086 434}%
\special{pa 4090 434}%
\special{pa 4096 434}%
\special{pa 4100 432}%
\special{pa 4106 430}%
\special{pa 4110 428}%
\special{pa 4116 424}%
\special{pa 4120 418}%
\special{pa 4126 414}%
\special{pa 4130 408}%
\special{pa 4136 404}%
\special{pa 4140 400}%
\special{pa 4146 394}%
\special{pa 4150 392}%
\special{pa 4156 388}%
\special{pa 4160 388}%
\special{pa 4166 386}%
\special{pa 4170 388}%
\special{pa 4176 388}%
\special{pa 4180 392}%
\special{pa 4186 394}%
\special{pa 4190 398}%
\special{pa 4196 404}%
\special{pa 4200 408}%
\special{pa 4206 414}%
\special{pa 4210 418}%
\special{pa 4216 424}%
\special{pa 4220 428}%
\special{pa 4226 430}%
\special{pa 4230 432}%
\special{pa 4236 434}%
\special{pa 4240 434}%
\special{pa 4246 434}%
\special{pa 4250 432}%
\special{pa 4256 430}%
\special{pa 4260 426}%
\special{pa 4266 422}%
\special{pa 4270 416}%
\special{pa 4276 412}%
\special{pa 4280 406}%
\special{pa 4286 402}%
\special{pa 4290 396}%
\special{pa 4296 394}%
\special{pa 4300 390}%
\special{pa 4306 388}%
\special{pa 4310 386}%
\special{pa 4316 386}%
\special{pa 4320 388}%
\special{pa 4326 390}%
\special{pa 4330 392}%
\special{sp}%
\special{pn 13}%
\special{pa 3390 410}%
\special{pa 3396 406}%
\special{pa 3400 400}%
\special{pa 3406 396}%
\special{pa 3410 392}%
\special{pa 3416 390}%
\special{pa 3420 388}%
\special{pa 3426 386}%
\special{pa 3430 386}%
\special{pa 3436 388}%
\special{pa 3440 390}%
\special{pa 3446 394}%
\special{pa 3450 398}%
\special{pa 3456 402}%
\special{pa 3460 406}%
\special{pa 3466 412}%
\special{pa 3470 416}%
\special{pa 3476 422}%
\special{pa 3480 426}%
\special{pa 3486 430}%
\special{pa 3490 432}%
\special{pa 3496 434}%
\special{pa 3500 434}%
\special{pa 3506 434}%
\special{pa 3510 432}%
\special{pa 3516 430}%
\special{pa 3520 426}%
\special{pa 3526 422}%
\special{pa 3530 418}%
\special{pa 3536 414}%
\special{pa 3540 408}%
\special{pa 3546 404}%
\special{pa 3550 398}%
\special{pa 3556 394}%
\special{pa 3560 390}%
\special{pa 3566 388}%
\special{pa 3570 386}%
\special{pa 3576 386}%
\special{pa 3580 388}%
\special{pa 3586 388}%
\special{pa 3590 392}%
\special{pa 3596 396}%
\special{pa 3600 400}%
\special{pa 3606 404}%
\special{pa 3610 408}%
\special{pa 3616 414}%
\special{pa 3620 420}%
\special{pa 3626 424}%
\special{pa 3630 428}%
\special{pa 3636 430}%
\special{pa 3640 432}%
\special{pa 3646 434}%
\special{pa 3650 434}%
\special{pa 3656 434}%
\special{pa 3660 432}%
\special{pa 3666 428}%
\special{pa 3670 426}%
\special{pa 3676 420}%
\special{pa 3680 416}%
\special{pa 3686 410}%
\special{pa 3690 406}%
\special{pa 3696 400}%
\special{pa 3700 396}%
\special{pa 3706 392}%
\special{pa 3710 390}%
\special{pa 3716 388}%
\special{pa 3720 386}%
\special{pa 3726 386}%
\special{pa 3730 388}%
\special{pa 3736 390}%
\special{pa 3740 394}%
\special{pa 3746 396}%
\special{pa 3750 402}%
\special{pa 3756 406}%
\special{pa 3760 412}%
\special{pa 3766 416}%
\special{pa 3770 422}%
\special{pa 3776 426}%
\special{pa 3780 430}%
\special{pa 3786 432}%
\special{pa 3790 434}%
\special{pa 3796 434}%
\special{pa 3800 434}%
\special{pa 3806 432}%
\special{pa 3810 430}%
\special{pa 3816 428}%
\special{pa 3820 424}%
\special{pa 3826 418}%
\special{pa 3830 414}%
\special{pa 3836 408}%
\special{pa 3840 404}%
\special{pa 3846 398}%
\special{pa 3850 394}%
\special{pa 3856 392}%
\special{pa 3860 388}%
\special{pa 3866 388}%
\special{pa 3870 386}%
\special{pa 3876 388}%
\special{pa 3880 388}%
\special{pa 3886 392}%
\special{pa 3890 394}%
\special{pa 3896 398}%
\special{pa 3900 404}%
\special{pa 3906 408}%
\special{pa 3910 414}%
\special{pa 3916 418}%
\special{pa 3920 424}%
\special{pa 3926 428}%
\special{pa 3930 430}%
\special{pa 3936 432}%
\special{pa 3940 434}%
\special{pa 3946 434}%
\special{pa 3950 434}%
\special{pa 3956 432}%
\special{pa 3960 428}%
\special{pa 3966 426}%
\special{pa 3970 420}%
\special{pa 3976 416}%
\special{pa 3980 412}%
\special{pa 3986 406}%
\special{pa 3990 402}%
\special{pa 3996 396}%
\special{pa 4000 392}%
\special{pa 4006 390}%
\special{pa 4010 388}%
\special{pa 4016 386}%
\special{pa 4020 386}%
\special{pa 4026 388}%
\special{pa 4030 390}%
\special{pa 4036 392}%
\special{pa 4040 396}%
\special{pa 4046 402}%
\special{pa 4050 406}%
\special{pa 4056 412}%
\special{pa 4060 416}%
\special{pa 4066 420}%
\special{pa 4070 426}%
\special{pa 4076 428}%
\special{pa 4080 432}%
\special{pa 4086 434}%
\special{pa 4090 434}%
\special{pa 4096 434}%
\special{pa 4100 432}%
\special{pa 4106 430}%
\special{pa 4110 428}%
\special{pa 4116 424}%
\special{pa 4120 418}%
\special{pa 4126 414}%
\special{pa 4130 408}%
\special{pa 4136 404}%
\special{pa 4140 400}%
\special{pa 4146 394}%
\special{pa 4150 392}%
\special{pa 4156 388}%
\special{pa 4160 388}%
\special{pa 4166 386}%
\special{pa 4170 388}%
\special{pa 4176 388}%
\special{pa 4180 392}%
\special{pa 4186 394}%
\special{pa 4190 398}%
\special{pa 4196 404}%
\special{pa 4200 408}%
\special{pa 4206 414}%
\special{pa 4210 418}%
\special{pa 4216 424}%
\special{pa 4220 428}%
\special{pa 4226 430}%
\special{pa 4230 432}%
\special{pa 4236 434}%
\special{pa 4240 434}%
\special{pa 4246 434}%
\special{pa 4250 432}%
\special{pa 4256 430}%
\special{pa 4260 426}%
\special{pa 4266 422}%
\special{pa 4270 416}%
\special{pa 4276 412}%
\special{pa 4280 406}%
\special{pa 4286 402}%
\special{pa 4290 396}%
\special{pa 4296 394}%
\special{pa 4300 390}%
\special{pa 4306 388}%
\special{pa 4310 386}%
\special{pa 4316 386}%
\special{pa 4320 388}%
\special{pa 4326 390}%
\special{pa 4330 392}%
\special{sp}%
\special{pn 13}%
\special{pa 3390 1010}%
\special{pa 3396 1006}%
\special{pa 3400 1000}%
\special{pa 3406 996}%
\special{pa 3410 992}%
\special{pa 3416 990}%
\special{pa 3420 988}%
\special{pa 3426 986}%
\special{pa 3430 986}%
\special{pa 3436 988}%
\special{pa 3440 990}%
\special{pa 3446 994}%
\special{pa 3450 998}%
\special{pa 3456 1002}%
\special{pa 3460 1006}%
\special{pa 3466 1012}%
\special{pa 3470 1016}%
\special{pa 3476 1022}%
\special{pa 3480 1026}%
\special{pa 3486 1030}%
\special{pa 3490 1032}%
\special{pa 3496 1034}%
\special{pa 3500 1034}%
\special{pa 3506 1034}%
\special{pa 3510 1032}%
\special{pa 3516 1030}%
\special{pa 3520 1026}%
\special{pa 3526 1022}%
\special{pa 3530 1018}%
\special{pa 3536 1014}%
\special{pa 3540 1008}%
\special{pa 3546 1004}%
\special{pa 3550 998}%
\special{pa 3556 994}%
\special{pa 3560 990}%
\special{pa 3566 988}%
\special{pa 3570 986}%
\special{pa 3576 986}%
\special{pa 3580 988}%
\special{pa 3586 988}%
\special{pa 3590 992}%
\special{pa 3596 996}%
\special{pa 3600 1000}%
\special{pa 3606 1004}%
\special{pa 3610 1008}%
\special{pa 3616 1014}%
\special{pa 3620 1020}%
\special{pa 3626 1024}%
\special{pa 3630 1028}%
\special{pa 3636 1030}%
\special{pa 3640 1032}%
\special{pa 3646 1034}%
\special{pa 3650 1034}%
\special{pa 3656 1034}%
\special{pa 3660 1032}%
\special{pa 3666 1028}%
\special{pa 3670 1026}%
\special{pa 3676 1020}%
\special{pa 3680 1016}%
\special{pa 3686 1010}%
\special{pa 3690 1006}%
\special{pa 3696 1000}%
\special{pa 3700 996}%
\special{pa 3706 992}%
\special{pa 3710 990}%
\special{pa 3716 988}%
\special{pa 3720 986}%
\special{pa 3726 986}%
\special{pa 3730 988}%
\special{pa 3736 990}%
\special{pa 3740 994}%
\special{pa 3746 996}%
\special{pa 3750 1002}%
\special{pa 3756 1006}%
\special{pa 3760 1012}%
\special{pa 3766 1016}%
\special{pa 3770 1022}%
\special{pa 3776 1026}%
\special{pa 3780 1030}%
\special{pa 3786 1032}%
\special{pa 3790 1034}%
\special{pa 3796 1034}%
\special{pa 3800 1034}%
\special{pa 3806 1032}%
\special{pa 3810 1030}%
\special{pa 3816 1028}%
\special{pa 3820 1024}%
\special{pa 3826 1018}%
\special{pa 3830 1014}%
\special{pa 3836 1008}%
\special{pa 3840 1004}%
\special{pa 3846 998}%
\special{pa 3850 994}%
\special{pa 3856 992}%
\special{pa 3860 988}%
\special{pa 3866 988}%
\special{pa 3870 986}%
\special{pa 3876 988}%
\special{pa 3880 988}%
\special{pa 3886 992}%
\special{pa 3890 994}%
\special{pa 3896 998}%
\special{pa 3900 1004}%
\special{pa 3906 1008}%
\special{pa 3910 1014}%
\special{pa 3916 1018}%
\special{pa 3920 1024}%
\special{pa 3926 1028}%
\special{pa 3930 1030}%
\special{pa 3936 1032}%
\special{pa 3940 1034}%
\special{pa 3946 1034}%
\special{pa 3950 1034}%
\special{pa 3956 1032}%
\special{pa 3960 1028}%
\special{pa 3966 1026}%
\special{pa 3970 1020}%
\special{pa 3976 1016}%
\special{pa 3980 1012}%
\special{pa 3986 1006}%
\special{pa 3990 1002}%
\special{pa 3996 996}%
\special{pa 4000 992}%
\special{pa 4006 990}%
\special{pa 4010 988}%
\special{pa 4016 986}%
\special{pa 4020 986}%
\special{pa 4026 988}%
\special{pa 4030 990}%
\special{pa 4036 992}%
\special{pa 4040 996}%
\special{pa 4046 1002}%
\special{pa 4050 1006}%
\special{pa 4056 1012}%
\special{pa 4060 1016}%
\special{pa 4066 1020}%
\special{pa 4070 1026}%
\special{pa 4076 1028}%
\special{pa 4080 1032}%
\special{pa 4086 1034}%
\special{pa 4090 1034}%
\special{pa 4096 1034}%
\special{pa 4100 1032}%
\special{pa 4106 1030}%
\special{pa 4110 1028}%
\special{pa 4116 1024}%
\special{pa 4120 1018}%
\special{pa 4126 1014}%
\special{pa 4130 1008}%
\special{pa 4136 1004}%
\special{pa 4140 1000}%
\special{pa 4146 994}%
\special{pa 4150 992}%
\special{pa 4156 988}%
\special{pa 4160 988}%
\special{pa 4166 986}%
\special{pa 4170 988}%
\special{pa 4176 988}%
\special{pa 4180 992}%
\special{pa 4186 994}%
\special{pa 4190 998}%
\special{pa 4196 1004}%
\special{pa 4200 1008}%
\special{pa 4206 1014}%
\special{pa 4210 1018}%
\special{pa 4216 1024}%
\special{pa 4220 1028}%
\special{pa 4226 1030}%
\special{pa 4230 1032}%
\special{pa 4236 1034}%
\special{pa 4240 1034}%
\special{pa 4246 1034}%
\special{pa 4250 1032}%
\special{pa 4256 1030}%
\special{pa 4260 1026}%
\special{pa 4266 1022}%
\special{pa 4270 1016}%
\special{pa 4276 1012}%
\special{pa 4280 1006}%
\special{pa 4286 1002}%
\special{pa 4290 996}%
\special{pa 4296 994}%
\special{pa 4300 990}%
\special{pa 4306 988}%
\special{pa 4310 986}%
\special{pa 4316 986}%
\special{pa 4320 988}%
\special{pa 4326 990}%
\special{pa 4330 992}%
\special{sp}%
%
\special{pn 8}%
\special{pa 4130 1000}%
\special{pa 4130 400}%
\special{fp}%
%
\special{pn 8}%
\special{pa 3380 400}%
\special{pa 3380 1000}%
\special{ip}%
%
\special{pn 8}%
\special{pa 1300 620}%
\special{pa 920 1000}%
\special{fp}%
\special{pa 1300 500}%
\special{pa 800 1000}%
\special{fp}%
\special{pa 1270 410}%
\special{pa 680 1000}%
\special{fp}%
\special{pa 1150 410}%
\special{pa 600 960}%
\special{fp}%
\special{pa 1030 410}%
\special{pa 600 840}%
\special{fp}%
\special{pa 910 410}%
\special{pa 600 720}%
\special{fp}%
\special{pa 790 410}%
\special{pa 600 600}%
\special{fp}%
\special{pa 670 410}%
\special{pa 600 480}%
\special{fp}%
\special{pa 1300 740}%
\special{pa 1040 1000}%
\special{fp}%
\special{pa 1300 860}%
\special{pa 1160 1000}%
\special{fp}%
%
\special{pn 4}%
\special{pa 2340 420}%
\special{pa 1780 980}%
\special{fp}%
\special{pa 2470 410}%
\special{pa 1900 980}%
\special{fp}%
\special{pa 2490 510}%
\special{pa 2020 980}%
\special{fp}%
\special{pa 2490 630}%
\special{pa 2110 1010}%
\special{fp}%
\special{pa 2490 750}%
\special{pa 2230 1010}%
\special{fp}%
\special{pa 2490 870}%
\special{pa 2360 1000}%
\special{fp}%
\special{pa 2210 430}%
\special{pa 1780 860}%
\special{fp}%
\special{pa 2090 430}%
\special{pa 1780 740}%
\special{fp}%
\special{pa 1970 430}%
\special{pa 1780 620}%
\special{fp}%
\special{pa 1870 410}%
\special{pa 1780 500}%
\special{fp}%
%
\special{pn 8}%
\special{pa 4130 490}%
\special{pa 3620 1000}%
\special{fp}%
\special{pa 4070 430}%
\special{pa 3480 1020}%
\special{fp}%
\special{pa 3950 430}%
\special{pa 3400 980}%
\special{fp}%
\special{pa 3870 390}%
\special{pa 3380 880}%
\special{fp}%
\special{pa 3730 410}%
\special{pa 3380 760}%
\special{fp}%
\special{pa 3600 420}%
\special{pa 3380 640}%
\special{fp}%
\special{pa 3470 430}%
\special{pa 3380 520}%
\special{fp}%
\special{pa 4130 610}%
\special{pa 3750 990}%
\special{fp}%
\special{pa 4130 730}%
\special{pa 3880 980}%
\special{fp}%
\special{pa 4130 850}%
\special{pa 4000 980}%
\special{fp}%
\special{pa 4130 970}%
\special{pa 4080 1020}%
\special{fp}%
%
\special{pn 4}%
\special{pa 5230 590}%
\special{pa 4820 1000}%
\special{fp}%
\special{pa 5230 470}%
\special{pa 4690 1010}%
\special{fp}%
\special{pa 5150 430}%
\special{pa 4600 980}%
\special{fp}%
\special{pa 5050 410}%
\special{pa 4520 940}%
\special{fp}%
\special{pa 4930 410}%
\special{pa 4520 820}%
\special{fp}%
\special{pa 4800 420}%
\special{pa 4520 700}%
\special{fp}%
\special{pa 4670 430}%
\special{pa 4520 580}%
\special{fp}%
\special{pa 5230 710}%
\special{pa 4950 990}%
\special{fp}%
\special{pa 5230 830}%
\special{pa 5080 980}%
\special{fp}%
\special{pa 5230 950}%
\special{pa 5200 980}%
\special{fp}%
\put(30.2000,-7.5000){\makebox(0,0)[lb]{$gg$}}%
\put(3.0000,-7.5000){\makebox(0,0)[lb]{$q\bar{q}$}}%
%
\special{pn 13}%
\special{pa 1420 410}%
\special{pa 1430 410}%
\special{fp}%
\special{sh 1}%
\special{pa 1430 410}%
\special{pa 1364 390}%
\special{pa 1378 410}%
\special{pa 1364 430}%
\special{pa 1430 410}%
\special{fp}%
%
\special{pn 13}%
\special{pa 1370 1010}%
\special{pa 1360 1010}%
\special{fp}%
\special{sh 1}%
\special{pa 1360 1010}%
\special{pa 1428 1030}%
\special{pa 1414 1010}%
\special{pa 1428 990}%
\special{pa 1360 1010}%
\special{fp}%
\put(14.3000,-14.0000){\makebox(0,0)[lb]{(a)}}%
\put(42.5000,-14.0000){\makebox(0,0)[lb]{(b)}}%
\end{picture}%

%% file: fig51.tex
\unitlength 0.1in
\begin{picture}( 18.4300, 11.1900)(  1.3000,-13.0700)
%
\special{pn 13}%
\special{pa 344 516}%
\special{pa 1134 516}%
\special{fp}%
\special{pa 1134 516}%
\special{pa 1926 200}%
\special{fp}%
\special{pa 344 992}%
\special{pa 1134 992}%
\special{fp}%
\special{pa 1134 992}%
\special{pa 1926 1308}%
\special{fp}%
\special{pa 1926 1308}%
\special{pa 1926 1308}%
\special{fp}%
%
\special{pn 13}%
\special{pa 344 834}%
\special{pa 976 834}%
\special{fp}%
\special{pa 976 834}%
\special{pa 1926 454}%
\special{fp}%
\special{pa 344 676}%
\special{pa 976 676}%
\special{fp}%
\special{pa 976 676}%
\special{pa 1134 738}%
\special{fp}%
\special{pa 1198 770}%
\special{pa 1926 1070}%
\special{fp}%
%
\special{pn 13}%
\special{pa 344 516}%
\special{pa 660 516}%
\special{fp}%
\special{sh 1}%
\special{pa 660 516}%
\special{pa 594 496}%
\special{pa 608 516}%
\special{pa 594 536}%
\special{pa 660 516}%
\special{fp}%
%
\special{pn 13}%
\special{pa 336 676}%
\special{pa 652 676}%
\special{fp}%
\special{sh 1}%
\special{pa 652 676}%
\special{pa 586 656}%
\special{pa 600 676}%
\special{pa 586 696}%
\special{pa 652 676}%
\special{fp}%
%
\special{pn 13}%
\special{pa 976 834}%
\special{pa 598 834}%
\special{fp}%
\special{sh 1}%
\special{pa 598 834}%
\special{pa 664 854}%
\special{pa 650 834}%
\special{pa 664 814}%
\special{pa 598 834}%
\special{fp}%
%
\special{pn 13}%
\special{pa 914 992}%
\special{pa 598 992}%
\special{fp}%
\special{sh 1}%
\special{pa 598 992}%
\special{pa 664 1012}%
\special{pa 650 992}%
\special{pa 664 972}%
\special{pa 598 992}%
\special{fp}%
%
\special{pn 13}%
\special{pa 1134 516}%
\special{pa 1530 358}%
\special{fp}%
\special{sh 1}%
\special{pa 1530 358}%
\special{pa 1462 364}%
\special{pa 1480 378}%
\special{pa 1476 402}%
\special{pa 1530 358}%
\special{fp}%
%
\special{pn 13}%
\special{pa 1198 770}%
\special{pa 1522 904}%
\special{fp}%
\special{sh 1}%
\special{pa 1522 904}%
\special{pa 1468 860}%
\special{pa 1474 884}%
\special{pa 1454 898}%
\special{pa 1522 904}%
\special{fp}%
%
\special{pn 13}%
\special{pa 1846 486}%
\special{pa 1490 628}%
\special{fp}%
\special{sh 1}%
\special{pa 1490 628}%
\special{pa 1560 622}%
\special{pa 1540 608}%
\special{pa 1546 584}%
\special{pa 1490 628}%
\special{fp}%
%
\special{pn 13}%
\special{pa 1824 1268}%
\special{pa 1484 1126}%
\special{fp}%
\special{sh 1}%
\special{pa 1484 1126}%
\special{pa 1538 1170}%
\special{pa 1532 1146}%
\special{pa 1552 1132}%
\special{pa 1484 1126}%
\special{fp}%
\put(1.3000,-8.4100){\makebox(0,0)[lb]{$N$}}%
\put(19.7300,-3.5800){\makebox(0,0)[lb]{$\phi$}}%
\put(19.7300,-12.5200){\makebox(0,0)[lb]{$\phi$}}%
\end{picture}%

%% file: fig52.tex
\unitlength 0.1in
\begin{picture}( 19.0800,  9.6600)(  2.2000,-11.4100)
%
\special{pn 13}%
\special{pa 374 542}%
\special{pa 1230 542}%
\special{fp}%
\special{pa 1230 542}%
\special{pa 2086 200}%
\special{fp}%
\special{pa 374 800}%
\special{pa 1230 800}%
\special{fp}%
\special{pa 1230 800}%
\special{pa 2086 1142}%
\special{fp}%
%
\special{pn 13}%
\special{pa 2086 928}%
\special{pa 1402 662}%
\special{fp}%
\special{pa 1402 662}%
\special{pa 2086 406}%
\special{fp}%
\put(21.2800,-3.4500){\makebox(0,0)[lb]{$\phi$}}%
\put(21.2800,-10.7300){\makebox(0,0)[lb]{$\phi$}}%
\put(2.2000,-7.3900){\makebox(0,0)[lb]{$N'$}}%
%
\special{pn 13}%
\special{pa 374 542}%
\special{pa 802 542}%
\special{fp}%
\special{sh 1}%
\special{pa 802 542}%
\special{pa 736 522}%
\special{pa 750 542}%
\special{pa 736 562}%
\special{pa 802 542}%
\special{fp}%
%
\special{pn 13}%
\special{pa 974 800}%
\special{pa 734 800}%
\special{fp}%
\special{sh 1}%
\special{pa 734 800}%
\special{pa 800 820}%
\special{pa 786 800}%
\special{pa 800 780}%
\special{pa 734 800}%
\special{fp}%
%
\special{pn 13}%
\special{pa 1230 542}%
\special{pa 1744 338}%
\special{fp}%
\special{sh 1}%
\special{pa 1744 338}%
\special{pa 1674 344}%
\special{pa 1694 358}%
\special{pa 1690 380}%
\special{pa 1744 338}%
\special{fp}%
%
\special{pn 13}%
\special{pa 2086 414}%
\special{pa 1778 518}%
\special{fp}%
\special{sh 1}%
\special{pa 1778 518}%
\special{pa 1848 516}%
\special{pa 1828 500}%
\special{pa 1834 478}%
\special{pa 1778 518}%
\special{fp}%
%
\special{pn 13}%
\special{pa 2086 1142}%
\special{pa 1700 988}%
\special{fp}%
\special{sh 1}%
\special{pa 1700 988}%
\special{pa 1754 1030}%
\special{pa 1750 1008}%
\special{pa 1770 994}%
\special{pa 1700 988}%
\special{fp}%
%
\special{pn 13}%
\special{pa 1530 706}%
\special{pa 1820 826}%
\special{fp}%
\special{sh 1}%
\special{pa 1820 826}%
\special{pa 1766 782}%
\special{pa 1772 806}%
\special{pa 1752 818}%
\special{pa 1820 826}%
\special{fp}%
\end{picture}%

%% file: fig53.tex
\unitlength 0.1in
\begin{picture}( 19.0600,  9.6900)(  2.3000,-11.4500)
%
\special{pn 13}%
\special{pa 2094 930}%
\special{pa 1406 664}%
\special{fp}%
\special{pa 1406 664}%
\special{pa 2094 406}%
\special{fp}%
\put(21.3600,-3.4600){\makebox(0,0)[lb]{$\phi$}}%
\put(21.3600,-10.7600){\makebox(0,0)[lb]{$\phi$}}%
%
\special{pn 13}%
\special{pa 1236 544}%
\special{pa 1750 338}%
\special{fp}%
\special{sh 1}%
\special{pa 1750 338}%
\special{pa 1682 344}%
\special{pa 1700 358}%
\special{pa 1696 380}%
\special{pa 1750 338}%
\special{fp}%
%
\special{pn 13}%
\special{pa 2094 416}%
\special{pa 1784 518}%
\special{fp}%
\special{sh 1}%
\special{pa 1784 518}%
\special{pa 1854 516}%
\special{pa 1836 502}%
\special{pa 1842 478}%
\special{pa 1784 518}%
\special{fp}%
%
\special{pn 13}%
\special{pa 2094 1146}%
\special{pa 1708 990}%
\special{fp}%
\special{sh 1}%
\special{pa 1708 990}%
\special{pa 1762 1034}%
\special{pa 1756 1010}%
\special{pa 1776 996}%
\special{pa 1708 990}%
\special{fp}%
%
\special{pn 13}%
\special{pa 1536 708}%
\special{pa 1828 828}%
\special{fp}%
\special{sh 1}%
\special{pa 1828 828}%
\special{pa 1774 784}%
\special{pa 1778 808}%
\special{pa 1758 820}%
\special{pa 1828 828}%
\special{fp}%
%
\special{pn 13}%
\special{pa 1236 544}%
\special{pa 2094 200}%
\special{fp}%
%
\special{pn 13}%
\special{pa 2094 1146}%
\special{pa 1236 802}%
\special{fp}%
%
\special{pn 13}%
\special{pa 1236 544}%
\special{pa 1236 802}%
\special{fp}%
\special{pn 13}%
\special{pa 376 544}%
\special{pa 380 538}%
\special{pa 386 530}%
\special{pa 390 526}%
\special{pa 396 524}%
\special{pa 400 524}%
\special{pa 406 526}%
\special{pa 410 530}%
\special{pa 416 538}%
\special{pa 420 544}%
\special{pa 426 552}%
\special{pa 430 558}%
\special{pa 436 562}%
\special{pa 440 564}%
\special{pa 446 562}%
\special{pa 450 560}%
\special{pa 456 554}%
\special{pa 460 546}%
\special{pa 466 540}%
\special{pa 470 532}%
\special{pa 476 526}%
\special{pa 480 524}%
\special{pa 486 524}%
\special{pa 490 526}%
\special{pa 496 530}%
\special{pa 500 536}%
\special{pa 506 544}%
\special{pa 510 550}%
\special{pa 516 558}%
\special{pa 520 562}%
\special{pa 526 564}%
\special{pa 530 564}%
\special{pa 536 560}%
\special{pa 540 554}%
\special{pa 546 548}%
\special{pa 550 540}%
\special{pa 556 534}%
\special{pa 560 528}%
\special{pa 566 524}%
\special{pa 570 522}%
\special{pa 576 524}%
\special{pa 580 528}%
\special{pa 586 534}%
\special{pa 590 542}%
\special{pa 596 550}%
\special{pa 600 556}%
\special{pa 606 562}%
\special{pa 610 564}%
\special{pa 616 564}%
\special{pa 620 562}%
\special{pa 626 556}%
\special{pa 630 550}%
\special{pa 636 542}%
\special{pa 640 534}%
\special{pa 646 528}%
\special{pa 650 524}%
\special{pa 656 522}%
\special{pa 660 524}%
\special{pa 666 528}%
\special{pa 670 534}%
\special{pa 676 540}%
\special{pa 680 548}%
\special{pa 686 554}%
\special{pa 690 560}%
\special{pa 696 564}%
\special{pa 700 564}%
\special{pa 706 562}%
\special{pa 710 558}%
\special{pa 716 550}%
\special{pa 720 544}%
\special{pa 726 536}%
\special{pa 730 530}%
\special{pa 736 526}%
\special{pa 740 524}%
\special{pa 746 524}%
\special{pa 750 526}%
\special{pa 756 532}%
\special{pa 760 540}%
\special{pa 766 546}%
\special{pa 770 554}%
\special{pa 776 560}%
\special{pa 780 562}%
\special{pa 786 564}%
\special{pa 790 562}%
\special{pa 796 558}%
\special{pa 800 552}%
\special{pa 806 544}%
\special{pa 810 538}%
\special{pa 816 530}%
\special{pa 820 526}%
\special{pa 826 524}%
\special{pa 830 524}%
\special{pa 836 526}%
\special{pa 840 530}%
\special{pa 846 538}%
\special{pa 850 544}%
\special{pa 856 552}%
\special{pa 860 558}%
\special{pa 866 562}%
\special{pa 870 564}%
\special{pa 876 562}%
\special{pa 880 560}%
\special{pa 886 554}%
\special{pa 890 546}%
\special{pa 896 540}%
\special{pa 900 532}%
\special{pa 906 526}%
\special{pa 910 524}%
\special{pa 916 524}%
\special{pa 920 526}%
\special{pa 926 530}%
\special{pa 930 536}%
\special{pa 936 544}%
\special{pa 940 550}%
\special{pa 946 558}%
\special{pa 950 562}%
\special{pa 956 564}%
\special{pa 960 564}%
\special{pa 966 560}%
\special{pa 970 554}%
\special{pa 976 548}%
\special{pa 980 540}%
\special{pa 986 534}%
\special{pa 990 528}%
\special{pa 996 524}%
\special{pa 1000 522}%
\special{pa 1006 524}%
\special{pa 1010 528}%
\special{pa 1016 534}%
\special{pa 1020 542}%
\special{pa 1026 550}%
\special{pa 1030 556}%
\special{pa 1036 562}%
\special{pa 1040 564}%
\special{pa 1046 564}%
\special{pa 1050 562}%
\special{pa 1056 556}%
\special{pa 1060 550}%
\special{pa 1066 542}%
\special{pa 1070 534}%
\special{pa 1076 528}%
\special{pa 1080 524}%
\special{pa 1086 522}%
\special{pa 1090 524}%
\special{pa 1096 528}%
\special{pa 1100 534}%
\special{pa 1106 540}%
\special{pa 1110 548}%
\special{pa 1116 554}%
\special{pa 1120 560}%
\special{pa 1126 564}%
\special{pa 1130 564}%
\special{pa 1136 562}%
\special{pa 1140 558}%
\special{pa 1146 550}%
\special{pa 1150 544}%
\special{pa 1156 536}%
\special{pa 1160 530}%
\special{pa 1166 526}%
\special{pa 1170 524}%
\special{pa 1176 524}%
\special{pa 1180 526}%
\special{pa 1186 532}%
\special{pa 1190 540}%
\special{pa 1196 546}%
\special{pa 1200 554}%
\special{pa 1206 560}%
\special{pa 1210 562}%
\special{pa 1216 564}%
\special{pa 1220 562}%
\special{pa 1226 558}%
\special{pa 1230 552}%
\special{pa 1236 544}%
\special{sp}%
\special{pn 13}%
\special{pa 376 804}%
\special{pa 380 796}%
\special{pa 386 788}%
\special{pa 390 784}%
\special{pa 396 782}%
\special{pa 400 782}%
\special{pa 406 784}%
\special{pa 410 788}%
\special{pa 416 796}%
\special{pa 420 804}%
\special{pa 426 810}%
\special{pa 430 816}%
\special{pa 436 820}%
\special{pa 440 822}%
\special{pa 446 822}%
\special{pa 450 818}%
\special{pa 456 812}%
\special{pa 460 804}%
\special{pa 466 798}%
\special{pa 470 790}%
\special{pa 476 784}%
\special{pa 480 782}%
\special{pa 486 780}%
\special{pa 490 784}%
\special{pa 496 788}%
\special{pa 500 794}%
\special{pa 506 802}%
\special{pa 510 808}%
\special{pa 516 816}%
\special{pa 520 820}%
\special{pa 526 822}%
\special{pa 530 822}%
\special{pa 536 818}%
\special{pa 540 812}%
\special{pa 546 806}%
\special{pa 550 798}%
\special{pa 556 792}%
\special{pa 560 786}%
\special{pa 566 782}%
\special{pa 570 780}%
\special{pa 576 782}%
\special{pa 580 786}%
\special{pa 586 792}%
\special{pa 590 800}%
\special{pa 596 808}%
\special{pa 600 814}%
\special{pa 606 820}%
\special{pa 610 822}%
\special{pa 616 822}%
\special{pa 620 820}%
\special{pa 626 814}%
\special{pa 630 808}%
\special{pa 636 800}%
\special{pa 640 792}%
\special{pa 646 786}%
\special{pa 650 782}%
\special{pa 656 780}%
\special{pa 660 782}%
\special{pa 666 786}%
\special{pa 670 792}%
\special{pa 676 798}%
\special{pa 680 806}%
\special{pa 686 812}%
\special{pa 690 818}%
\special{pa 696 822}%
\special{pa 700 822}%
\special{pa 706 820}%
\special{pa 710 816}%
\special{pa 716 808}%
\special{pa 720 802}%
\special{pa 726 794}%
\special{pa 730 788}%
\special{pa 736 784}%
\special{pa 740 780}%
\special{pa 746 782}%
\special{pa 750 784}%
\special{pa 756 790}%
\special{pa 760 798}%
\special{pa 766 804}%
\special{pa 770 812}%
\special{pa 776 818}%
\special{pa 780 822}%
\special{pa 786 822}%
\special{pa 790 820}%
\special{pa 796 816}%
\special{pa 800 810}%
\special{pa 806 804}%
\special{pa 810 796}%
\special{pa 816 788}%
\special{pa 820 784}%
\special{pa 826 782}%
\special{pa 830 782}%
\special{pa 836 784}%
\special{pa 840 788}%
\special{pa 846 796}%
\special{pa 850 804}%
\special{pa 856 810}%
\special{pa 860 816}%
\special{pa 866 820}%
\special{pa 870 822}%
\special{pa 876 822}%
\special{pa 880 818}%
\special{pa 886 812}%
\special{pa 890 804}%
\special{pa 896 798}%
\special{pa 900 790}%
\special{pa 906 784}%
\special{pa 910 782}%
\special{pa 916 780}%
\special{pa 920 784}%
\special{pa 926 788}%
\special{pa 930 794}%
\special{pa 936 802}%
\special{pa 940 808}%
\special{pa 946 816}%
\special{pa 950 820}%
\special{pa 956 822}%
\special{pa 960 822}%
\special{pa 966 818}%
\special{pa 970 812}%
\special{pa 976 806}%
\special{pa 980 798}%
\special{pa 986 792}%
\special{pa 990 786}%
\special{pa 996 782}%
\special{pa 1000 780}%
\special{pa 1006 782}%
\special{pa 1010 786}%
\special{pa 1016 792}%
\special{pa 1020 800}%
\special{pa 1026 808}%
\special{pa 1030 814}%
\special{pa 1036 820}%
\special{pa 1040 822}%
\special{pa 1046 822}%
\special{pa 1050 820}%
\special{pa 1056 814}%
\special{pa 1060 808}%
\special{pa 1066 800}%
\special{pa 1070 792}%
\special{pa 1076 786}%
\special{pa 1080 782}%
\special{pa 1086 780}%
\special{pa 1090 782}%
\special{pa 1096 786}%
\special{pa 1100 792}%
\special{pa 1106 798}%
\special{pa 1110 806}%
\special{pa 1116 812}%
\special{pa 1120 818}%
\special{pa 1126 822}%
\special{pa 1130 822}%
\special{pa 1136 820}%
\special{pa 1140 816}%
\special{pa 1146 808}%
\special{pa 1150 802}%
\special{pa 1156 794}%
\special{pa 1160 788}%
\special{pa 1166 784}%
\special{pa 1170 780}%
\special{pa 1176 782}%
\special{pa 1180 784}%
\special{pa 1186 790}%
\special{pa 1190 798}%
\special{pa 1196 804}%
\special{pa 1200 812}%
\special{pa 1206 818}%
\special{pa 1210 822}%
\special{pa 1216 822}%
\special{pa 1220 820}%
\special{pa 1226 816}%
\special{pa 1230 810}%
\special{pa 1236 804}%
\special{sp}%
\put(2.3000,-7.3200){\makebox(0,0)[lb]{$G$}}%
\end{picture}%